\documentstyle[11pt,aaspp4,psfig]{article}

\lefthead{Chiappini et al.}
\righthead{Formation of the Milky Way}

\begin{document}

\title{Abundance Gradients and the Formation of the Milky Way}

\author{Cristina Chiappini\altaffilmark{1,2}}
\author{Francesca Matteucci\altaffilmark{3,4}}
\author{Donatella Romano\altaffilmark{4}}

\altaffiltext{1}{Osservatorio Astronomico di Trieste, 
		 Via G.B. Tiepolo 11, I-34131 Trieste, Italy;
                 chiappin@ts.astro.it}
\altaffiltext{2}{Department of Astronomy, Columbia University,
                 Mail Code 5247, Pupin Hall, 550 West 120th Street,
                 New York, NY 10027; chiappini@astro.columbia.edu}
\altaffiltext{3}{Dipartimento di Astronomia, Universit\`a di Trieste,
                 Via G.B. Tiepolo 11, I-34131 Trieste, Italy; 
		 matteucci@ts.astro.it}
\altaffiltext{4}{SISSA/ISAS, Via Beirut 2-4, I-34014 Trieste, Italy; 
       		 romano@sissa.it}




\begin{abstract}

In this paper we adopt a chemical evolution model, which is an improved 
version of the Chiappini, Matteucci, \& Gratton (1997) model, assuming two 
main accretion episodes for the formation of the Galaxy: the first one 
forming the halo and bulge in a short timescale followed by a second one that 
forms the thin-disk, with a timescale which is an increasing function of the 
Galactocentric distance (being of the order of 7 Gyrs at the solar 
neighborhood). The present model takes into account in more detail than 
previously the halo density distribution and explores the effects of a 
threshold density in the star formation process, during both the halo and 
disk phases. The model also includes the most recent nucleosynthesis 
prescriptions concerning supernovae of all types, novae and single stars 
dying as white dwarfs. In the comparison between model predictions and 
available data, we have focused our attention on abundance gradients as 
well as gas, stellar and star formation rate distributions along the disk, 
since this kind of model has already proven to be quite successful in 
reproducing the solar neighborhood characteristics. We suggest that the 
mechanism for the formation of the halo leaves detectable imprints on the 
chemical properties of the outer regions of the disk, whereas the 
evolution of the halo and the inner disk are almost completely disentangled.
This is due to the fact that the halo and disk 
densities are comparable at large Galactocentric distances and therefore the 
gas lost from the halo can substantially contribute to building  up the outer 
disk. We also show that the existence of a threshold density for the star 
formation rate, both in the halo and disk phase, is necessary to reproduce 
the majority of observational data in the solar vicinity and in the whole 
disk. In particular, a threshold in the star formation implies the occurrence 
of a gap in the star formation at the halo-disk transition phase, in 
agreement with recent data.

We conclude that  a relatively short halo formation timescale ($\simeq$ 0.8 
Gyr), in agreement with recent estimates for the age differences among 
Galactic globular clusters, coupled with an ``inside-out'' formation of the 
Galactic disk, where the innermost regions are assumed to have formed much 
faster than the outermost ones, represents, at the moment, the most likely 
explanation for the formation of the Milky Way. This scenario allows us to 
predict abundance gradients and other radial properties of the Galactic disk 
in very good agreement with observations. Moreover, as a consequence of the 
adopted ``inside-out'' scenario for the disk, we predict that the abundance 
gradients along the Galactic disk must have increased with time and that the 
average $<[\alpha/$Fe$]>$ ratios in stars (halo plus disk) slightly decrease 
going from 4 to 10 kpcs from the Galactic centre. We also show that the same 
ratios increase substantially towards the outermost disk regions and the 
expected scatter in the stellar ages decreases, because the outermost regions 
are dominated by halo stars. 
More observations at large Galactocentric distances 
are needed to test these predictions.
\end{abstract}




%

\section{INTRODUCTION}

Most of the progress in understanding the formation and evolution of the Milky 
Way galaxy comes, on one side, from observations concerning chemical abundances 
in stars and gas and from the continuous improvement of stellar nucleosynthesis 
calculations on the other side. Other important quantities relevant to the chemical 
evolution of the Milky Way, such as the star formation rate (SFR), the initial 
mass function (IMF) and gas flows, are still poorly constrained. However, a 
good model of chemical evolution can allow us to impose constraints on 
such quantities. In particular, a good chemical evolution model should be able 
to reproduce observables larger in number than the number of adopted 
free parameters. Among the observables there are some which can help us to 
constrain the model better than others (see Matteucci 
2000). The best observables from the point of view of Galactic chemical 
evolution are: the abundances in stars, in H\,{\small II} 
regions and planetary nebulae (PNe) from which we derive the abundance 
gradients along the Galactic disk, and the G-dwarf metallicity distribution 
in the solar neighborhood.

In past years a great deal of theoretical work has appeared concerning the 
chemical evolution of the Milky Way (Matteucci \& Fran\c cois 1989; Carigi 
1994, 1996; Timmes et al. 1995; Tsujimoto et al. 1995; Chiappini et al. 1997 
(hereafter CMG97), 1999; Chang et al. 1999; Portinari \& Chiosi 1999, 2000; 
Boissier \& Prantzos 1999; Goswami \& Prantzos 2000; Romano et al. 2000) 
and most of them dealt with the evolution of the Galactic disk only. Only 
a few models have attempted to model the Galactic halo (CMG97; Samland et 
al. 1997; Goswami \& Prantzos 2000). In particular, CMG97 have suggested a 
scenario where the Galaxy forms as a result of two main infall episodes. 
During the first episode the halo forms and the gas lost by the halo rapidly 
(roughly 1 Gyr) accumulates in the center with the consequent formation of the 
bulge. During the second episode, a much slower infall of primordial gas gives 
rise to the disk with the gas accumulating faster in the inner than in the 
outer regions. In this scenario the formation of the halo and disk are almost 
completely dissociated although some halo gas falls into the disk. This 
mechanism for disk formation is known as ``inside-out'' scenario which 
is quite successful in reproducing the main features of the Milky Way 
(CMG97) as well as of external galaxies especially concerning abundance 
gradients (see Prantzos \& Boissier, 2000). 

In this paper we present a model similar to that of CMG97 and Romano et al. 
(2000) but with an improved treatment of the halo evolution. The model is 
always based on the ``two-infall'' scenario but more attention is paid to 
the evolution of the halo and to the possibility of a density threshold for 
the SFR in the halo. In the CMG97 paper we did not investigate in detail the 
effects of a density threshold for the star formation in the halo as our main 
goal was to study the solar neighborhood. Moreover, as will be shown in the 
next sections, the evolution of the solar neighborhood is almost independent 
of the enrichment history of the halo.

It is worth recalling that a surface density threshold for star formation has 
been observed in a variety of objects, including normal spirals, starburst 
galaxies and low surface brightness galaxies (e.g Kennicutt 1989, 1998; van 
der Hulst et al. 1993). The existence of a threshold in spheroidal systems 
(halos, bulges and ellipticals) is not yet clear although there are 
theoretical arguments by Elmegreen (1999) suggesting the existence of a 
threshold also in such systems. The threshold adopted by CMG97 produced 
results in very good agreement with the majority of the observational 
constraints in the solar neighborhood and in the whole disk. These include the 
most recent findings which suggest a discontinuity in the chemical properties 
between the halo (and part of the thick-disk) and the thin-disk (e.g. Pagel 
\& Tautvaisiene 1995; Beers \& Sommer-Larsen 1995; Gratton et al. 1996). 
In particular, one of the main effects of the threshold was that it naturally 
produced a hiatus in the SFR between the end of the halo/thick-disk phase and 
the beginning of the thin-disk phase. Such a hiatus seems to be real since it 
is observed both in the plot of [Fe/O] versus [O/H] (CMG97; Gratton et al. 
1996, 2000; see also Pagel 2000 for a discussion) and in the plot of [Fe/Mg] 
versus [Mg/H] (Fuhrmann 1998; Gratton et al. 2000). The evidence for this is 
shown by the steep increase of [Fe/O] and [Fe/Mg] at a particular value of 
[O/H] and [Mg/H], respectively, indicating that at a certain epoch (coinciding 
with the halo-disk transition) SNe II, responsible for the production of O 
and Mg, stopped exploding while Fe, produced by the long living SNe Ia, 
continued to be produced. The star formation must have ceased for a period 
which cannot be longer than $\sim$ 1 Gyr, as suggested by theoretical models 
(CMG97; Gratton et al. 2000 and the present paper). The ``two-infall'' model 
also allowed us to fit the observed metallicity distribution of the G-dwarfs 
by assuming a long timescale for the thin-disk formation. This long timescale 
for the thin-disk formation at the solar vicinity was then suggested also by 
more recent chemical evolution models (e.g. Prantzos \& Silk 1998; Portinari, 
Chiosi, \& Bressan 1998; Chang et al. 1999; and more recently Boissier \& 
Prantzos 1999; Hou et al. 2000). The same result is also suggested by 
chemodynamical models (Samland et al. 1997; Hensler 1999).

We not that in CMG97 we focused mainly on the solar vicinity 
modelling, and adopted very simplified assumptions to compute the radial 
profiles. We predicted non-linear gradients along the Galactic disk and a 
tendency for them to be flatter in the outermost regions and almost constant 
with time, whereas in the innermost parts they were steeper and 
steepened with time. In particular, the gradients we obtained were flatter 
than the standard values presently adopted in the literature. For example, the 
gradient of oxygen was flatter than $-$0.07 dex/kpc, which is the value 
suggested by different data sources (B stars, H\,{\small II} regions and PNe, 
see Maciel \& Quireza 1999).

In this paper, we study in detail the formation of abundance gradients in the 
Galactic disk by trying to fit at the same time the abundance, SFR, star and 
gas profiles in the Milky Way (MW) disk. Clearly the evolution of gradients 
with time is strictly related to the mechanism of Galaxy formation, including 
the formation of the halo and this is an important point to assess. For this 
purpose, we test several assumptions about the halo and disk formation and the 
existence and variation of a density threshold for the SFR both in the halo 
and disk. Our results concern mostly the elements for which more data are 
available, such as N, O, S and Fe.

We believe that by improving the MW model we will provide a solid basis for a 
more detailed understanding of the history of chemical enrichment of our 
Galaxy and other spirals. An important aspect of this work will be to provide 
specific predictions for galaxy sizes of MW-like galaxies as a function of 
redshift type (in a forthcoming paper), which could in principle be tested 
against the growing body of observational data on distant galaxies. 
In particular, the slow timescale for the formation of the MW disk implied by 
our chemical evolution model implies that at high redshifts we should see 
smaller disks. Some observational tests, using the predictions of CMG97 with 
the growth of the thin-disk timescale as a function of Galactocentric 
distance, have already been done (Roche et al. 1998). Those authors estimate 
that galaxies at redshift 2 $<$ $z$ $<$ 3.5 show 2.79 $\pm$ 0.31 mag of surface 
brightness evolution relative to those at $z$ $<$ 0.35, which is significantly 
larger than the luminosity evolution over this redshift range. They suggest 
that this can be explained by a size and luminosity evolution model, in which 
the outer regions of spiral galaxies form later and with a longer timescale 
than the inner regions, causing the half-light radius to increase with time. 
Although the interpretation of the observational data is still controversial 
(see Simard et al. 1999), they clearly represent a powerful test for disk 
formation models.

The paper is organized as follows: in section 2 the observational constraints 
are discussed; section 3 presents the model assumptions and in section 4 the 
results are shown. Some conclusions are drawn in section 5.

\section{OBSERVATIONAL DATA}

\subsection{Abundance gradients}

Radial abundance profiles constitute one of the most important observational 
constraints for models of the evolution of the MW disk. However, over the past 
decade, different authors using different observational tools often came to 
contradictory views on both the shape, the magnitude and the evolution of the 
abundance gradients along the disk (see Tosi 1996, 2000 for a discussion on 
this point). The controversial results originate from both theoretical and 
observational considerations.

In particular, observations of B-stars seemed to rule out the existence of a 
gradient (or to allow for only a very mild one; Gehren et al. 1985; 
Fitzsimmons et al. 1992; Kaufer et al. 1994; Kilian-Montenbruck et al. 1994), 
in contrast to observations of H\,{\small II} regions (Shaver et al. 1983; 
Fich \& Silkey 1991; Simpson et al. 1995; Afflerbach et al. 1997; Rudolph et 
al. 1997) and PNe of type II (Maciel \& Chiappini 1994; Maciel \& K\"oppen 
1994; Maciel \& Quireza 1999). This controversy was 
solved by Smartt \& Rolleston (1997) and Gummersbach et al. (1998) who found, 
by means of the reanalysis of a sample of B-stars, a gradient similar to that 
derived from H\,{\small II} regions and PNe. 
 
Although the B-star discrepancy seems now to be solved, a new one appeared 
recently in the literature concerning the H\,{\small II} regions abundance 
gradients. Deharveng et al. (2000) analysed a new sample of 34 H\,{\small II} 
regions located between 6.6 and 17.7 kpcs from the Galactic center and after a 
careful estimate of the electron temperatures in those objects they 
obtained, using their best observations, an oxygen abundance gradient which is 
flatter (by a factor of 2) than the one obtained in previous works based on 
H\,{\small II} regions. Their result seems to be in good agreement with 
results by Esteban et al. (1998, 1999a, b). Proposals for flatter gradients 
or bimodal ones have also been made by works based on open clusters (Friel 
1999 and Twarog et al. 1997) although the situation is still very unclear (see 
Carraro et al. 1998). Observations of open clusters seem to support the 
existence of an iron gradient along the disk, but large uncertainties related 
to both the metallicity calibration and the ages of these objects are still 
present. 

Another open question related to the observed abundance gradients concerns 
their variation with time: do the gradients steepen or flatten with time? This 
question cannot be answered properly by the presently available data (see 
Maciel 2000 for a discussion). However, PNe are the most promising objects to 
solve this problem. As has been extensively discussed in the past few years, 
PNe Galactic distribution, kinematics, chemical composition and morphology 
clearly indicate that PNe comprise objects of different populations (Peimbert 
1978; Maciel 1997). Previous work has shown that disk objects of type II are 
particularly useful in tracing the chemical enrichment of the interstellar 
medium at the time of the formation of the PN progenitors (Maciel \& K\"oppen 
1994; Maciel \& Chiappini 1994).

In a recent work, Maciel \& Quireza (1999) obtained the gradients for a sample 
that includes the objects studied by Maciel \& K\"oppen (1994), Maciel \& 
Chiappini (1994) and Costa et al. (1997) (the latter consists of a sample of 
PNe near the anticentre direction intended to derive a better estimate of the 
gradients at Galactocentric distances larger than the solar position). Their 
main conclusions were: {\it i)} there is an average gradient of $-$0.04 to 
$-$0.07 dex/kpc for what they call ``inner'' Galaxy (between 4 and 10 kpc 
--- those authors assume $R_\odot$ = 7.6 kpc); {\it ii)} the gradients show 
a small variation for the different element ratios (see their table 2); {\it 
iii)} the PN gradients are generally slightly flatter than those derived from 
younger objects; {\it iv)} for larger Galactocentric distances the PN 
gradients show some flattening. However the authors pointed out that {\it a)} 
the precise region where the gradient flattens out is not well defined and 
larger samples are needed, particularly at $R$ $>$ 12 kpc; and {\it b)} 
although the tendency of the gradients to steepen with time suggested by 
Maciel \& K\"oppen (1994) seems to be confirmed by the more recent analysis, 
this is only an indicative result because of the difficulty of assigning 
precise ages to disk PNe and of estimating the importance of dynamical effects 
which could have flattened the gradients of the older objects.

From the theoretical point of view, several mechanisms have been proposed in 
order to explain the existence of abundance gradients in disk galaxies (see 
Matteucci \& Chiappini 1999). Among the alternatives, the so called 
``biased-infall'' is the most popular one. In this approach the Galactic disk 
formed by infall of gas occurring at a faster rate in the innermost regions 
than in the outermost ones (``inside-out'' formation picture). The physical 
reason for a ``biased-infall'' can be found in the fact that the gas tends to 
collapse more quickly in the center of the spheroid so that a gradient in the 
gas density, lower in the more external parts, is established. In this 
situation the gas which continues to infall towards the disk falls more 
efficiently towards the center than towards the external regions, owing to the 
stronger tidal force acting at the center (see Larson 1976). Several authors 
have shown that the ``biased-infall'' can well reproduce the steep abundance 
gradients along the disk, especially if coupled with a SFR proportional to 
some power $k$ of the gas density (see section 3) with $k$ $>$ 1 (Tosi 1988; 
Matteucci \& Fran\c cois 1989).

However, the situation is still not clear also from a theoretical 
point of view. In fact, as pointed out in many recent papers 
(Maciel \& Quireza 1999; Tosi 2000; Portinari \& Chiosi 2000; Deharveng et al. 
2000; Hou et al. 2000 and others), different authors can fit the solar 
vicinity constraints and even the present time abundance gradients, but they 
seem to be divided into two groups concerning the evolution of the 
abundance gradients: in some of them the gradients steepen with time (Tosi 
1988; CMG97; Samland et al. 1997) while in others the abundance gradients 
flatten with time (Allen et al. 1998; Portinari \& Chiosi 2000; Hou et al. 
2000). The lack of good data for the {\it outer} MW disk abundances still 
prevents us from testing this prediction and represents one of the main 
reasons for the non-uniqueness of the various chemical evolution models. 
Table 1 summarizes the abundance data available to trace the gradients in our 
Galaxy. In this work we present results for the Galactic abundance gradients of 
N, O, S and Fe as those are the elements for which more data exist.

\subsection{Radial profiles}

\subsubsection{Gas}

The observed (H {\scriptsize I} + H$_2$) distribution is taken from Dame 
(1993). The surface density distribution of the total gas $\Sigma_{gas}$ 
is obtained from the sum of the H {\scriptsize I} and H$_2$ distributions, 
$\Sigma_{HI} + \Sigma_{H_2}$, accounting for the helium and heavy elements 
fractions (thick line in figure 4).

\subsubsection{Stars}

The stellar profile is exponentially decreasing outwards, with characteristic 
scale length $R_{stars}$ $\sim$ 2.5\,--\,3 kpc (Sackett 1997 and references 
therein; Freudenreich 1998). Moreover, COBE observations suggest that the 
stellar disk has an outer edge 4 kpc from the Sun (Freudenreich 1998). To 
compare our model predictions on the stellar density profile along the 
Galactic disk to the observed one we consider $\Sigma_{stars}(R_\odot, 
t_{Gal})$ = 35 $\pm$ 5 $M_\odot$ pc$^{-2}$ (Gilmore et al. 1989) and 
$R_{stars}$ = 2.5 kpc (see figure 5).

\subsubsection{Star formation rate}

The distributions of supernova remnants (Guibert, Lequeux, \& Viallefond 
1978), of pulsars (Lyne, Manchester, \& Taylor 1985), of Lyman-continuum 
photons (G\"usten \& Mezger 1982) and of molecular gas (Rana 1991 and 
references therein) all can be used to derive an estimate for the SFR along 
the Galactic disk. Since these observables cannot directly yield the absolute 
SFR without further assumptions --- e.g. on the IMF and mass ranges for 
producing pulsars, supernovae, etc. (see Lacey \& Fall 1985), it is common 
practice to normalize them to their values at the solar radius, and then to 
trace the radial profile SFR($R$)/SFR($R_\odot$) (the thick lines plotted in 
figure 6 refer to the upper and lower limits obtained from the observational 
data listed above).

\section{THE CHEMICAL EVOLUTION MODEL}

As we stated in the introduction, a good model for the chemical evolution of 
our Galaxy should honour a minimal number of observational constraints both 
in the solar neighborhood and in the whole disk. The model we will adopt here 
is an updated version of the so called ``two-infall'' model of CMG97.

This model assumes that the Galaxy forms out of two main accretion episodes 
almost completely dissociated. During the first episode, the primordial gas 
collapses very quickly and forms the spheroidal components (halo and bulge); 
during the second episode, the thin-disk forms, mainly by accretion of a 
significant fraction of matter of primordial chemical composition plus traces 
of halo gas. The disk is built-up in the framework of the ``inside-out'' 
scenario of Galaxy formation (Larson 1976) which ensures the formation of 
abundance gradients along the disk (Matteucci \& Fran\c cois 1989). 

As in Matteucci \& Fran\c cois (1989), the Galactic disk is approximated by 
several independent rings, 2 kpc wide, without exchange of matter between 
them. The rate of accretion of matter in each shell is:
\begin{equation}
\frac{d\Sigma_I(R, t)}{dt} = A(R)\,e^{- t/\tau_{H}} + B(R)\,e^{- (t - 
t_{max})/\tau_{D}},
\end{equation}
where $\Sigma_I(R, t)$ is the surface mass density of the infalling material, 
which is assumed to have primordial chemical composition; $t_{max}$ is the 
time of maximum gas accretion onto the disk, coincident with the end of the 
halo/thick-disk phase and set here equal to 1 Gyr; $\tau_{H}$ and 
$\tau_{D}$ are the timescales for mass accretion onto the halo/thick-disk and 
thin-disk components, respectively. In particular, $\tau_{H}$ = 0.8 Gyr and, 
according to the ``inside-out'' scenario, $\tau_{D}(R)$ = 1.033$\times 
(R$/kpc)$-$1.267 Gyr (see Romano et al. 2000). We adopt a linear approximation 
for the variation of $\tau_{D}(R)$. However, this will be best constrained by 
the observed radial profiles of gas, SFR and abundances. In some cases the 
observed radial profiles suggest that this timescale can approach a constant 
value for Galactocentric distances larger than $\simeq$ 12 kpc (see next 
section). The quantities $A(R)$ and $B(R)$ are derived from the condition of 
reproducing the current total mass surface density distribution in the halo 
and along the disk respectively at the present time. 

The SFR adopted here has the same formulation as in CMG97: 

\begin{equation}
\psi(R, t) = \nu(t)\,\left( \frac{\Sigma(R, t)}{\Sigma(R_\odot, t)} 
\right) ^{2\,(k - 1)}\,\left( \frac{\Sigma(R, t_{Gal})}{\Sigma(R, t)} \right)
^{k - 1}\,\Sigma_{gas}^{k}(R, t),
\end{equation}
where $\nu(t)$ is the efficiency of the star formation process, $\Sigma(R,t)$ 
is the total mass surface density at a given radius $R$ and given time $t$, 
$\Sigma(R_\odot, t)$ is the total mass surface density at the solar position 
and $\Sigma_{gas}(R, t)$ is the gas surface density. Note that the gas surface 
density exponent, $k$, equal to 1.5, was obtained from the best model of CMG97 
in order to ensure a good fit to the observational constraints at the solar 
vicinity. This value is also in very good agreement with the recent 
observational results of Kennicutt (1998) and with N-body simulation results 
by Gerritsen \& Icke (1997).

The efficiency of star formation is set to $\nu$=1 Gyr$^{-1}$ to ensure the 
best fit to the observational features in the solar vicinity, and becomes zero 
when the gas surface density drops below a certain critical threshold 
(Kennicutt 2000). We adopt a threshold density 
$\Sigma_{th}$ $\sim$ 7 $M_\odot$ pc$^{-2}$ 
in the disk (CMG97). As far as the halo/thick-disk phase is concerned, a 
similar value for the threshold is expected (see Elmegreen 1999 and 
the Model A description below for details). 

The IMF is that of Scalo (1986), assumed to be constant during the evolution 
of the Galaxy (see Chiappini, Matteucci \& Padoan 2000). Our present model 
differs from that of CMG97 in {\it (i)} the adopted yields for the low and 
intermediate mass range stars which are now taken from van den Hoek \& 
Groenewegen (1997) instead of Renzini \& Voli (1981) (see Romano et al. 2000); 
{\it (ii)} the fact that now we are including the explosive nucleosynthesis 
from nova outbursts (see Romano \& Matteucci 2000) 
and {\it (iii)} the adopted solar 
Galactocentric distance ($R_\odot$ = 8 $\pm$ 0.5 kpc, Reid 1993) and age of 
the Galaxy ($t_{Gal}$ = 14 Gyr). Moreover, in the present model we adopt a 
primordial $^4$He abundance of 0.241 (by mass) instead of 0.23 as recently 
suggested by Viegas, Gruenwald, \& Steigman (2000). The primordial abundances 
by mass of D and $^3$He were taken to be 4.5 $\times$ 10$^{-5}$ and 2.0 
$\times$ 10$^{-5}$ respectively (see Chiappini \& Matteucci 2000).
\par
We run several models by varying the SFR threshold in the halo phase, the 
surface mass density distribution of the halo and the timescale for the halo 
formation. The present surface mass density distribution of the disk has 
been always kept the same and is exponential with scale length $R_D$ = 3.5 kpc 
normalized to $\Sigma_D(R_\odot, t_{Gal})$ = 54 $M_\odot$ pc$^{-2}$ (see 
Romano et al. 2000 for a discussion of the choice of these parameters).

The models are labelled A, B, C, D and the model parameters are summarized 
in Table 5.

In particular, Model A assumes: 
\par
\par\noindent{$\bullet$} the total halo mass density profile is constant and 
			 equal to 17 $M_\odot$ pc$^{-2}$ for $R$ $\le$ 8 kpc 
			 and decreases as $R^{-1}$ outwards. We adopt the halo 
			 mass profile given by Binney \& Tremaine (1987) where 
			 the density is proportional to $R^{-2}$ and this 
			 implies a surface mass density profile proportional 
			 to  $R^{-1}$; the value of the total surface mass 
			 density of the halo at the solar position is quite 
			 uncertain, and we assumed it to be 17 $M_\odot$ pc$^{-2}$ 
			 from the fact that the total surface mass density 
			 (halo plus disk) in the solar vicinity is $\simeq$ 71 
			 $M_\odot$ pc$^{-2}$ (Kuijken and Gilmore 1991), where  
			 $\simeq$ 54 $M_\odot$ pc$^{-2}$ corresponds to the 
			 disk total surface mass density (see also a 
			 discussion in Romano et al. 2000). 

\medskip
\par\noindent{$\bullet$} $\tau_{H}$ is constant and equal to 0.8 Gyr along the 
			 overall Galactic disk, and $\tau_{D}(R)$ is an 
			 increasing function of $R$ in the range 4\,--\,14 kpc,
                         as described above, and constant and equal to 
			 $\tau_{D}$(14 kpc) for $R$ $>$ 14 kpc.
\medskip
\par\noindent{$\bullet$} the thresholds in the gas density are 4 and 7 
			 $M_\odot$ pc$^{-2}$ during the halo and disk phase, 
			 respectively. These choices are not arbitrary since 
			 the threshold for the disk agrees with the 
			 observational values found by Kennicutt (1989, 1998). 
			 Not much is known about the threshold in the halo and so
			 this particular choice is a result of several tests and
			 ensures a good fit to the main observational 
			 constraints. In particular, these choices 
			 for the density thresholds lead to a very good fit of 
			 the gap in the SFR during the halo-disk transition 
			 phase, as suggested by both the [Fe/O] vs. [O/H] and 
			 [Fe/Mg] vs. [Mg/H] relations, as well as of the 
			 radial profiles in the disk. 
\par
A possible physical motivation for different star formation
thresholds in the halo and the disk might be related to a metallicity
effect. Since the opacities are lower in the halo due to a lower metal content,
this would make it easier for a protostellar cloud to cool and collapse
and in this case we would expect a lower threshold density. 
Moreover, if the threshold density
is close to the tidal limit (which is the average total density inside
a radius R), then it would be higher in the inner regions than in the 
outer ones. Consequently one would expect a higher threshold density in 
the bulge than in the disk outside the bulge and higher in the disk than 
in the remote halo that is further out from the center (Elmegreen 1999).
However, it is worth nothing that in the present model we adopt a simplification
of the problem as both the disk and halo threshold density are taken
as constant with the galactocentric distance. A more physical treatment 
of the threshold (varying with position) will be pursued in
a forthcoming paper.
  
Model B assumes that:
\par
\par\noindent{$\bullet$} same as Model A, but considering a constant halo 
			 surface mass density of $\Sigma_H(R, t_{Gal})$ = 17 
			 $M_\odot$ pc$^{-2}$ at all Galactocentric distances.

Model C assumes that:
\par
\par\noindent {$\bullet$} same as Model A, but no threshold in the gas density 
	      during the halo/thick-disk phase.
\bigskip

Finally, Model D assumes that:
\medskip
\par\noindent {$\bullet$} same as Model A, but also $\tau_{H}$ varies as a 
	      function of $R$: $\tau_{H}(R)$ = 0.8 Gyr for $R$ $\le$ 10 Gyr, 
	      $\tau_{H}(R)$ = 2 Gyr for $R$ $\ge$ 12 Gyr. This slower 
	      formation of the outer halo might represent a merging scenario 
	      for the halo formation as originally suggested by Searle \& Zinn 
	      (1978). In other words this simulates an ``inside-out'' 
	      formation also for the halo. The situation, in this case, is 
	      similar to that predicted by models which do not take the halo 
	      into account in studying the formation of the disk (e.g. 
     	      Boissier \& Prantzos 1999; Portinari \& Chiosi 1999, 2000). 
	      These models predict steep abundance gradients in the outermost 
	      disk regions owing to the fact that they do not take into account 
	      the pre-enrichment of the disk due to the halo, which is not 
	      negligible at large Galactocentric distances.

\section{MODEL RESULTS}

\subsection{The solar vicinity}

All the above models can fit the available constraints in the solar vicinity 
(see table 2), however they differ substantially in reproducing the 
properties of the whole disk. In particular, concerning the solar 
neighborhood, our models are in good agreement with what is called the minimum 
set of observational constraints, among which the most important is the 
G-dwarf metallicity distribution (see Tosi 2000 for a recent review; see 
figure 1 for Model A). In table 3 we present the values for the observables in 
the solar vicinity predicted by Models A and C (the results from Models B and 
D are the same as Model A for the solar vicinity and they differ only for $R$ 
$>$ 8 kpc). In table 4 we show the solar abundances predicted by Model A 
compared to the observed ones (Anders \& Grevesse 1989; Grevesse \& Sauval 
1998) and the agreement is quite good. Model C gives almost the same 
predictions as Model A concerning solar abundances, so we do not list them 
here. This is because {\it at the solar radius} the chemical evolution of the 
thin-disk is not influenced by that of the halo/thick-disk. On the 
contrary, the abundances observed at the outermost radii strongly reflect the 
chemical enrichment occurring during the halo phase, as we will see in the next 
section. 

It is worth noting that all of these models provide a good fit of the 
[$\alpha$/Fe] versus [Fe/H] relation in the solar vicinity, as shown in 
Chiappini et al. (1999). Here we only show the plot of [Fe/O] versus [O/H] 
which indicates the existence of a gap in the SFR occurring at [O/H] $\sim$ 
$-$0.2 dex (see figure 2a). In fact, if there is a gap in the SFR we should 
expect both a steep increase of [Fe/O] at a fixed [O/H] and a lack of stars 
corresponding to the gap period. This is what both our model (Model A) and the 
data show. This gap, suggested also by the [Fe/Mg] data of Fuhrmann (1998), in 
our models is due to the adoption of the threshold in the star formation 
process coupled with the assumption of a slow infall for the formation of
the disk. This is evident from figure 3 which shows clearly a halt in the SFR 
at the halo-disk transition.  
A similar explanation for the gap in the star formation rate
is that it might result from the fact that the star formation in the halo 
begins while the protogalaxy is still in an overall expansion phase. 
Consequently the disk can only form later as the protogalaxy collapses 
and virializes (see Mathews and Schramm 1993).

It is worth noting here, that the data in figure 
2a belong to a large compilation presented in Gratton et al. (2000). These 
data contain abundances obtained both from permitted and forbidden lines and 
do not show the linear decrease in [Fe/O] (or the growth in [O/Fe]) for low 
metallicity exhibited by recent data from Israelian et al. (1998) and 
Boesgaard et al. (1999) obtained from UV OH lines. This oxygen
abundance controversy is 
particularly embarassing from a theoretical point of view, since a steep 
increase of [O/Fe] at low metallicities would imply the same behaviour also 
for the other $\alpha$-elements (but see Ramaty et al. 2000 for a solution 
involving only oxygen), and this, it seems, is not observed. 

If such behaviour is confirmed also for the other $\alpha$- elements 
(sharing a common origin with oxygen) we should revise the yields of Fe and/or 
those of oxygen. In fact, a linear increase of the [O/Fe] could be due either 
to a lower contribution of Fe from type II SNe or to a metallicity dependent 
yield of oxygen or, as a last possibility, to a strong variation of the 
IMF during the halo phase. An earlier appearance of type Ia SNe during the 
halo phase (at around [Fe/H] = $-$3.0) is unrealistic since it would imply 
that stars more massive than 8 $M_{\odot}$ can be progenitors of such 
supernovae. In fact, most of the models in the literature (including the 
present one) do not produce the observed linear behaviour of the [Fe/O] found 
by Israelian et al. (1998), although those models 
assume an 8 $M_{\odot}$ star as the 
maximum mass allowed to form a CO white dwarf, corresponding to a timescale of 
only about 300 Myr for the first explosions. However, it is worth noting that 
although the first type Ia SNe may occur quite early, the bulk of them occurs 
only after $\sim $ 1 Gyr (Matteucci \& Greggio 1986) and produce a visible 
change in the abundance ratios for metallicities larger than [Fe/H] $\simeq$ 
$-$1. Recent SN Ia models with maximum allowed progenitors smaller than 8 
$M_{\odot}$ have been suggested, but the chemical evolution results seem to be 
unaffected (Kobayashi et al. 1999), thus confirming that only the SN Ia 
originating from low mass stars have an impact on the abundances of the ISM. 
The small increase of [O/Fe] for [Fe/H] $<$ $-1$ is 
far from being the observed trend by Israelian et al. (1998), contrary to what 
is stated by Goswami \& Prantzos (2000). As explained in Chiappini et al. (1999) 
the reason why the so-called ``[O/Fe]-plateau'' is not exactly flat is due to 
a) the fact that some contribution by SN Ia can happen already after about 300 
Myr as explained above and b) the fact that the stellar evolution calculations 
for massive stars (Woosley \& Weaver 1995; Thielemann et al. 1996) predict 
that the ejected mass of (O/Fe)$_{ej}$ by a massive star is an increasing 
function of the stellar mass (see figure 2b).

\subsection{Radial profiles: does the halo evolution modify the abundance 
	    gradients?}

Our results for the radial profiles of gas, stars and SFR are shown in figures 
4, 5 and 6, respectively. It can be seen that Model A is the model in best 
agreement with the observations. Model B (constant halo density profile) 
overestimates the star density in the outer regions while Model D 
underestimates this quantity. Moreover, Model B gives a constant gas profile 
for Galactocentric distances between 8\,--\,20 kpc, at variance with 
observations.

In figures 7, 8 and 9 we compare the model predictions with the abundance data 
for N, O and S, respectively. In the upper panels the present time gradient is
shown: our model predictions are compared with abundance data obtained from 
H\,{\small II} regions and B-stars (see section 2). In the lower panels a 
comparison is made with data from PNe of type II (roughly 2 Gyrs old objects). 
In figure 10 the abundance data on open clusters are compared with our model 
predictions for the present day abundance gradient. We did not attempt to 
separate the open cluster sample by age.

Among all the gradient figures, Model A seems again to be the one which best 
agrees with the observations. Model B predicts gradients that are flatter than 
the observed ones (as in CMG97), while Model D predicts too steep gradients.

Model C is instead an alternative model which could be supported by the 
observations. In fact, given the scatter of the observations for a given 
Galactocentric distance, it is very difficult to constrain our models, 
especially in the outer parts of the disk. Model C assumes no density 
threshold for the SFR in the halo phase. As a consequence, a higher initial 
level of enrichment due to the halo evolution is reached in the Galactic disk, 
which is almost constant for Galactocentric distances larger than 12 kpc (see 
also the almost constant stellar profile predicted by this model in figure 5). 
This is due to the predominance of the halo gas in the outer regions of the 
disk and it creates a pre-enriched disk gas dominating the subsequently 
accreted primordial one.

Figure 11 shows the behaviour of the abundance gradients, for each element and 
each model, at four different epochs of the evolution of the Galaxy. In 
figure 12 we show the abundance gradient behaviour as a function of 
time (between 4 and 14 kpcs). As is clearly shown by this figure the inner 
gradients tend to steepen with time while the outer gradients remain 
essentially constant. This conclusion is in agreement with CMG97 and with the 
chemodynamical model results (Samland et al. 1997). This behaviour is expected 
in a model taking into account the ``inside-out'' formation of the disk, where 
the external disk regions are still forming now and the abundance gradient is 
still building up. In fact, at early epochs, the efficiency in the chemical 
enrichment of the inner Galactic regions is low (due to the large amount of 
primordial infalling gas) leading to a flat initial abundance gradient. Then, 
at late epochs, while the SFR is still much higher in the central than in the 
external regions the infall of metal poor gas is stronger in the outer than in 
the inner regions, thus steepening the gradients.

However, other authors find a flattening of the gradients (see Tosi 2000 for a 
clear discussion on the possible scenarios for the evolution of the abundance 
gradients). A flattening of the gradients with time can clearly be achieved if 
one assumes that the disk formed on the same timescale at any Galactocentric 
distance, but with this scenario it is very difficult to reproduce the correct 
gradients at the present time. However, models adopting the 
``inside-out'' picture for the disk formation also seem to create gradients 
flattening with time. This can be explained as follows:  in the inner parts of 
the disk those models assume a very 
high efficiency in the chemical enrichment process 
already in the earliest phases of the Galaxy evolution, thus soon reaching a 
maximum metallicity in the gas which then remains constant or decreases due to 
the gas recycled by dying stars (Portinari \& Chiosi 1999; Hou et al. 2000;
Boissier \& Prantzos 1999). At the same time, in the outermost disk regions 
the lack of any pre-enrichment from the halo phase (a similar situation is 
shown by our model D discussed above) and the fact that those models 
do not include 
a threshold in the star formation process in the disk, produces a growth of 
metallicity which we do not find. This also explains an important difference 
between the results shown in our figure 11 and those of figure 4 of Hou et al. 
(2000). As in their model the halo phase is completely decoupled from the disk 
evolution (even in the outer parts), their initial metallicities at larger 
Galactocentric distances are very small. Moreover, since they do not consider 
a threshold on the star formation process in the disk, their predicted 
abundances in the outermost parts of the disk keep increasing with time 
leading to a flattening of the abundance gradients.

In summary, from figure 12 we can conclude that:

\medskip
\par\noindent {\it a)} Model B gives the flattest gradients. 
		 This model is
		 in good agreement with the recent flatter gradient suggested 
		 by Deharveng et al. (2000). The predicted gas and stellar 
		 density profiles are at variance with the observed ones 
                 in the 
		 outer parts of the disk. However, we stress that large
		 uncertainties are still present in the observed profiles 
		 especially at larger Galactocentric distances.

\par\noindent {\it b)} Model C predicts a present time
		  gradient of O in agreement with the 
		 ``standard'' adopted value of $-$0.07 to $-$0.06 dex/kpc as
		 suggested by B-stars, PNe and most of the data based on 
		 H\,{\small II} regions. Morever, it predicts a steeper 
		 gradient for Fe and N than for O and S, which is in agreement 
		 for instance with the results by Shaver et al. (1983) (see 
		 table 1) where a gradient of $-$0.07 dex/kpc is measured for 
		 oxygen while a value of $-$0.09 dex/kpc is obtained for N.
                 The fact that the N gradient is steeper than the O gradient 
		 is not surprising since N is mainly a secondary element (i.e.
                 its abundance is metallicity dependent) and is mainly 
                 produced 
                 on long timescales as is iron.

\par\noindent {\it c)} Model A also predicts gradients that steepen with time 
		 but only in the first 5 Gyrs of the Galactic disk evolution, 
		 remaining essentially constant with time after that. Again 
		 the gradients obtained for O and S are flatter than the ones 
		 obtained for N and Fe. The abundance gradients obtained in 
		 Model A for the present time are slightly flatter than the 
		 ones of Model C (for a Galactocentric range of 4\,--\,14 kpc).

Finally, in figure 13 we show the predicted $<$[O/Fe]$>$ ratios in the stars 
born and still alive at any Galactocentric distance (including halo, 
thick- and thin-disk stars) as a function of such distance. In particular, 
this ratio refers to the stellar population dominating in mass at any radius. 
It is clear from the figure that, while the ratios are slightly decreasing 
in the Galactocentric distance range 4\,--\,10 kpcs, as expected from the 
``inside-out'' disk formation, they substantially increase at large radii. 
This is due to the fact that for $R > 10$ kpcs the halo/thick-disk stars are 
dominating and they have large [O/Fe] ratios. In the same figure we show also 
the plot of the same ratios versus the stellar age of the dominating stellar 
population and this confirms that the stars dominating in the outermost disk 
regions are the oldest ones. {\it This also means that the spread in stellar 
ages is a decreasing function of the Galactocentric distance}. Edvardsson 
(1998) showed similar plots but only for disk stars older than 10 Gyrs and 
located between 4 and 14 kpcs and, although a precise comparison cannot be 
made, they seem to be in agreement with our predictions.

\medskip

\subsubsection{Radial profiles: main results}

\bigskip
\par\noindent{$\bullet$} The outer gradients are sensitive to the halo 
			 evolution, in particular to the amount of halo 
gas which ends up in the disk. This result is not surprising since the 
halo density is comparable to that of the outer disk, whereas it is negligible 
when
compared to that of the inner disk. 
Therefore, the inner parts of the disk
                         ($R$ $<$ $R_\odot$) evolve  
			 independently from the halo evolution in agreement 
			 with CMG97 result.

\medskip
\par\noindent{$\bullet$} Our best-model predicts gradients in good 
			 agreement with the observed ones in PNe, H\,{\small 
			 II} regions and open clusters. In CMG97 we 
                         found flatter 
			 gradients than in the present paper
and in Matteucci \& Fran\c cois (1989),
although a very similar ``inside-out'' scenario was assumed in all these papers.
The reason for being flatter than in Matteucci \& Fran\c cois (1989)
results from the fact that in CMG97 we adopted a 
threshold density for the SFR. This means that every time the gas density goes 
below the threshold then the star formation stops and starts again only after 
the gas has reached a density above the threshold, because 
of new infalling gas 
and/or the gas restored by dying stars (see figure 3). In particular, the 
effect 
of such a threshold was to keep the gas density always close to the threshold 
value especially at late times and large radii, 
thus contributing to maintaining an almost 
constant metallicity irrespective of the Galactocentric distance. 
The reason for the fact that the present model predicts steeper gradients than
CMG97, although a threshold is also assumed, is that we considered here 
a more realistic halo density distribution, namely, a decreasing function
of the galactocentric distance for $R >$ 8kpc, leading to a smaller contribution
of the halo to the disk component metallicity enrichment at large radii. 

\par\noindent{$\bullet$} We predict that the abundance gradients along the 
			 Galactic disk must have increased with time. 
This is a direct consequence of the assumed ``inside-out'' scenario for the 
formation of the Galactic disk. Moreover, the gradients of different elements 
are predicted to be slightly different, owing to their different nucleosynthesis 
histories. In particular, Fe and N, which are produced on longer timescales than 
the $\alpha$-elements, show steeper gradients. Unfortunately, the available 
observations cannot yet confirm or disprove this, because the predicted 
differences are below the limit of detectability.

\medskip
\par\noindent{$\bullet$} Our model demonstrates  
                         a satisfactory fit to 
		         the elemental abundance gradients and it is also in 
			 good agreement with the observed radial profiles of 
			 the SFR, gas density and the number of stars in the 
			 disk.

\medskip
\par\noindent{$\bullet$} Our best model suggests that the average 
$<[\alpha/$Fe]$>$ ratio in stars slightly decreases from 4 to 10 kpcs.
This is due to the predominance of disk over halo stars in this distance range 
and to the fact that the ``inside-out'' scenario for the disk predicts a decrease of 
such ratios (Matteucci 1991). On the other hand we predict a substantial 
increase ($\sim 0.3$ dex) of these ratios in the range 10\,--\,18 kpcs, 
owing to the predominance, in this region, of the halo over the disk stars.

\section{CONCLUSIONS}

In this paper we have presented an improved version of the model of CMG97
for the chemical evolution of the Milky Way. This
model assumes that the halo and disk form almost independently out of two 
different episodes of infall of gas of primordial chemical composition.
This scenario is suggested by many observational facts and seems the most 
likely at the present time, even though our understanding of the formation and 
evolution of the Milky Way is still far from clear.
We explored several cases where we varied in turn the timescale for the 
formation of the halo, which is always much smaller than for the disk, the halo
density distribution as a function of Galactocentric distance and the density 
threshold for the SFR in the halo phase. 
The existence of such a threshold is suggested both by observational and 
theoretical arguments (Kennicutt 1989; Elmegreen 1999). 
Concerning the disk we have assumed an ``inside-out'' formation scenario, where 
the timescale for disk formation is a linear function of the Galactocentric 
distance. This assumption has proven to be successful in explaining the main 
features of the Galactic disk as well as those of other spirals.
The novelty of our approach consists in the fact that we have found that the 
halo formation strongly affects the evolution of the
outermost regions of the disk whereas it leaves untouched the innermost ones. We 
have considered mostly the radial properties of the disk, such as abundance 
gradients, star, gas and SFR distributions. The evolution of the 
solar neighborhood is unchanged relative to our previous papers (CMG97 and 
Chiappini et al. 1999).
Our model did not consider radial flows (a thorough discussion of the effects 
of radial flows can be found in Portinari \& Chiosi, 2000).
In any case, radial flows by themselves cannot be the main cause of gradients 
along the disk although their presence can help in building up gradients, but 
always under very specific conditions.

Our general conclusions can be summarized as follows: 

\begin{itemize}
\item[-] A decoupling between halo and disk phases is needed in order to best 
	 fit all the observational constraints. However, the halo 
	 evolution can affect the outer regions of the disk evolution given the 
	 fact that at larger Galactocentric distances the disk is still in 
	 process of forming and has very low gas densities and metallicities.

\item[-] The threshold in the star formation process is important not only as 
	 a factor shaping the abundance gradients but also because it 
	 naturally produces a star formation gap between the halo and the 
	 disk phases (in agreement with some recent abundance results, e.g. 
	 Fuhrmann 1998; Gratton et al. 2000). Moreover our results indicate
         that the threshold gas density in the (outer) halo is lower than
         the one in the disk.

\item[-] Long timescale for the disk formation:  
	 $\tau_D(R_\odot$ = 8 
	 kpc) = 7 Gyrs and longer times at larger radii are needed 
(in agreement with CMG97). This has been 
	 confirmed also by recent papers (Portinari et al. 1998; Boissier \& 
	 Prantzos 1999).

\item[-] Better agreeement with the observational constraints 
         is obtained for a constant rather than variable IMF 
         (Chiappini et al. 
	 2000) and this conclusion is mostly based on the abundance gradients
         and radial profiles of gas and SFR.


\item[-] We predict abundance gradients along the Galactic disk in very good 
agreement with observations (although there is still some controversy about the
absolute value of such gradients) and gradients of different elements are sligthly 
different according to their nucleosynthetic origin. 
In particular, we predict that although there is a slight flattening
of the abundance gradients at intermediate galactocentric distances,
they are very steep in the outermost regions of the disk.
Unfortunately, there are no available data for abundance gradients in the
outermost disk regions, but a study of the metallicity in the gas in 
the extreme outer regions of
disks of spirals 
(Ferguson et al. 1998), has revealed that there is no evidence for 
a flattening of the gradients.  
We also predict that abundance gradients have increased with time. 

\item[-] We predict a slight decrease with distance in the average $<[\alpha$/Fe]$>$ 
ratios in stars born in the Galactocentric distance range 4\,--\,10 Kpcs and
an increase with distance of this ratio in the range 10\,--\,18 Kpc. 
In addition, a smaller spread of stellar ages
with increasing Galactocentric distance
is found.

\item[-] We conclude that a scenario where the halo (mostly the inner one)
formed relatively quickly ($\sim$ 0.8 Gyr), in agreement with recent estimates 
for the age differences among Galactic globular clusters (Rosenberg et al. 
1999), 
and the disk grew differentially (``inside-out''),
represents the most likely explanation for the formation and 
evolution of the Milky Way.

\item[-] However, in order to draw more secure conclusions 
it is necessary to have, in 
the future, more data in the outermost regions of the disk!
\end{itemize}

\begin{acknowledgements}
One of us (F.M.) thanks the Institute of Astronomy in Cambridge 
(U.K.) for its
kind hospitality. Support from the Italian Ministry for University and
Scientific and Technological Research (MURST) is gratefully acknowledged. 
C.C. wish to thank Thomas M. Dame for having kindly sent his data on the gas
density distribution along the disk and E. Friel for having sent us
her data on iron abundances in open clusters in advance of publication.
C.C. acknowledges financial support from Columbia University.
The authors also thank Dr. John Danziger and the referee Dr. Grant Mathews 
for their comments and suggestions.

\end{acknowledgements}

\vfill\eject

\footnotesize
\begin{table}
\centering
\small
\textsc{Table 1. Summary of abundance gradients.}
\vspace{0.3cm}
\footnotesize
\label{}
\begin{tabular}{c c c c c c}
\hline
Element & Study & Gradient & R$_\odot$ & Radial Baseline & Tool\\
        &     &  (dex kpc$^{-1}$) &  (kpc) & (kpc) & \\
\hline
Oxygen & Shaver et al. 1983 & $-$0.07 $\pm$ 0.015 & 10 & 5---13 & 
H\,{\scriptsize II} regions\\
       & Gehren et al. 1985 & $-$0.01 $\pm$ 0.02 &  & 8---18 & 
B-type stars\\
       & Fitzsimmons et al. 1992 & $-$0.03 $\pm$ 0.02 & 8.5 & 6---13 &
B-type stars\\
       & Kaufer et al. 1994 & $-$0.000 $\pm$ 0.009 & 8.5 & 6---17 &
B-type stars\\
       & Kilian-Montenbruck et al. 1994 & $-$0.021 $\pm$ 0.012 & 8.7 &
6---15 & B-type stars\\
       & Maciel \& K\"oppen 1994 & $-$0.069 $\pm$ 0.006 & 8.5 & 4---13 &
Planetary nebulae\\
       & V\' \i lchez \& Esteban 1996 & $-$0.036 $\pm$ 0.020 & 8.5 & 12---18 &
H\,{\scriptsize II} regions\\
       & Afflerbach et al. 1997 & $-$0.064 $\pm$ 0.009 & 8.5 & 0---12 &
H\,{\scriptsize II} regions\\
       & Smartt \& Rolleston 1997 & $-$0.07 $\pm$ 0.01 & 8.5 & 6---18 &
B-type stars\\
       & Gummersbach et al. 1998 & $-$0.067 $\pm$ 0.024 & 8.5 & 5---14 &
B-type stars\\
       & Maciel \& Quireza 1999 & $-$0.058 $\pm$ 0.007 & 7.6 & 3---14 &
Planetary nebulae\\
       & Deharveng et al. 2000 & $-$0.039 $\pm$ 0.005 & 8.5 & 5---15 &
H\,{\scriptsize II} regions\\
Nitrogen & Shaver et al. 1983 & $-$0.09 $\pm$ 0.015 & 10 & 5---13 & 
H\,{\scriptsize II} regions\\
         & Kaufer et al. 1994 & $-$0.026 $\pm$ 0.009 & 8.5 & 6---17 &
B-type stars\\
         & Kilian-Montenbruck et al. 1994 & $-$0.017 $\pm$ 0.020 & 8.7 &
6---15 & B-type stars\\
         & Simpson et al. 1995 & $-$0.10 $\pm$ 0.02 & 8.5 & 0---10 &
H\,{\scriptsize II} regions\\
         & V\' \i lchez \& Esteban 1996 & $-$0.009 $\pm$ 0.020 & 8.5 & 
12---18 & H\,{\scriptsize II} regions\\
         & Afflerbach et al. 1997 & $-$0.072 $\pm$ 0.006 & 8.5 & 0---12 &
H\,{\scriptsize II} regions\\
         & Rudolph et al. 1997 & $-$0.111 $\pm$ 0.012 & 8.5 & 0---17 &
H\,{\scriptsize II} regions\\
         & Gummersbach et al. 1998 & $-$0.078 $\pm$ 0.023 & 8.5 & 5---14 &
B-type stars\\
Sulphur & Shaver et al. 1983 & $-$0.01 $\pm$ 0.020 & 10 & 5---13 & 
H\,{\scriptsize II} regions\\
        & Kilian-Montenbruck et al. 1994 & $-$0.026 $\pm$ 0.025 & 8.7 &
6---15 & B-type stars\\
        & Maciel \& K\"oppen 1994 & $-$0.067 $\pm$ 0.006 & 8.5 & 4---13 &
Planetary nebulae\\
        & Simpson et al. 1995 & $-$0.07 $\pm$ 0.02 & 8.5 & 0---10 &
H\,{\scriptsize II} regions\\
        & V\' \i lchez \& Esteban 1996 & $-$0.041 $\pm$ 0.020 & 8.5 & 
12---18 & H\,{\scriptsize II} regions\\
        & Afflerbach et al. 1997 & $-$0.063 $\pm$ 0.006 & 8.5 & 0---12 &
H\,{\scriptsize II} regions\\
        & Rudolph et al. 1997 & $-$0.079 $\pm$ 0.009 & 8.5 & 0---17 &
H\,{\scriptsize II} regions\\
        & Maciel \& Quireza 1999 & $-$0.077 $\pm$ 0.011 & 7.6 & 3---14 &
Planetary nebulae\\
Neon & Kilian-Montenbruck et al. 1994 & $-$0.043 $\pm$ 0.011 & 8.7 &
6---15 & B-type stars\\
     & Maciel \& K\"oppen 1994 & $-$0.056 $\pm$ 0.007 & 8.5 & 4---13 &
Planetary nebulae\\
     & Simpson et al. 1995 & $-$0.08 $\pm$ 0.02 & 8.5 & 0---10 &
H\,{\scriptsize II} regions\\
     & Maciel \& Quireza 1999 & $-$0.036 $\pm$ 0.010 & 7.6 & 3---14 &
Planetary nebulae\\
Iron & Friel \& Janes 1993 & $-$0.09 $\pm$ 0.02 & 8.5 & 7---16 &
Open clusters\\
     & Kilian-Montenbruck et al. 1994 & $-$0.003 $\pm$ 0.020 & 8.7 &
6---15 & B-type stars\\
     & Twarog et al. 1997 & $-$0.067 $\pm$ 0.008 & 8.5 & 6---16 & 
Open clusters\\
     &   & $-$0.023 $\pm$ 0.017 & 8.5 & 6---10 & Open clusters\\
     &   & $-$0.004 $\pm$ 0.018 & 8.5 & 10---16 & Open clusters\\
     & Carraro et al. 1998 & $-$0.09 & 8.5 & 7---16 & Open clusters\\
     & Friel 1999 & $-$0.06 $\pm$ 0.01 & 8.5 & 7---16 & Open clusters\\
Helium & Shaver et al. 1983 & $-$0.001 $\pm$ 0.008 & 10 & 5---13 & 
H\,{\scriptsize II} regions\\
Carbon & Kilian-Montenbruck et al. 1994 & $+$0.001 $\pm$ 0.015 & 8.7 &
6---15 & B-type stars\\
       & Gummersbach et al. 1998 & $-$0.035 $\pm$ 0.014 & 8.5 & 5---14 &
B-type stars\\
Magnesium & Kilian-Montenbruck et al. 1994 & $-$0.020 $\pm$ 0.011 & 8.7 &
6---15 & B-type stars\\
          & Gummersbach et al. 1998 & $-$0.082 $\pm$ 0.026 & 8.5 & 5---14 &
B-type stars\\
Silicon & Kilian-Montenbruck et al. 1994 & $+$0.000 $\pm$ 0.018 & 8.7 &
6---15 & B-type stars\\
        & Gummersbach et al. 1998 & $-$0.107 $\pm$ 0.028 & 8.5 & 5---14 &
B-type stars\\
\hline
\end{tabular}
\end{table}

\vfill\eject

\begin{table}
\centering
\small
\textsc{Table 2. Main observational constraints for the solar neighborhood.}
\vspace{0.5cm}
\footnotesize
\label{}
\begin{tabular}{c c c}
\hline
Observable & Observed value & Reference\\
\hline
Surface densities of: & & \\
gas & 13 $\pm$ 3 $M_\odot$ pc$^{-2}$ & Kulkarni \& Heiles 1987\\
    & 7 $M_\odot$ pc$^{-2}$ & Dickey 1993\\
stars (alive) & 35 $\pm$ 5 $M_\odot$ pc$^{-2}$ & Gilmore et al. 1995\\
stars (WDs + NSs) & 2---4 $M_\odot$ pc$^{-2}$ & M\'era et al. 1998\\
total & 48 $\pm$ 9 $M_\odot$ pc$^{-2}$ & Kuijken \& Gilmore 1991\\
      & 84$^{+30}_{-25}$ $M_\odot$ pc$^{-2}$ & Bahcall et al. 1992\\
      & 52 $\pm$ 13 $M_\odot$ pc$^{-2}$ & Flynn \& Fuchs 1994\\
      & 50---60 $M_\odot$ pc$^{-2}$ & Cr\'ez\'e et al. 1998; Holmberg \& 
	Flynn 1998\\
Star formation rate & 2---10 $M_\odot$ pc$^{-2}$ Gyr$^{-1}$ & G\"usten \& 
Mezger 1982\\
SNI rate & 0.3 $\pm$ 0.2 century$^{-1}$ & Cappellaro \& Turatto 1996\\
SNII rate & 1.2 $\pm$ 0.8 century$^{-1}$ & Cappellaro \& Turatto 1996\\
Nova rate & 20---30 yr$^{-1}$ & Shafter 1997\\
Infall rate & 0.3---1.5 $M_\odot$ pc$^{-2}$ Gyr$^{-1}$ & Portinari et al. 
1998\\
Metal-poor/total stars & 2---10 \% & Pagel \& Patchett 1975; Matteucci et al.
1990\\
\hline
\end{tabular}
\end{table}

\begin{table}
\centering
\small
\textsc{Table 3. Observed and predicted quantities at $R_\odot$ and 
	$t$ = $t_{Gal}$.}
\vspace{0.5cm}
\footnotesize
\label{}
\begin{tabular}{c c c c}
\hline
  & A & C & Observed\\
\hline
Metal-poor/total stars (\%)  &  10 \% &  10 \% & 2---10 \% \\
SNIa (century$^{-1}$) & 0.4 & 0.4 & 0.3 $\pm$ 0.2 \\
SNII (century$^{-1}$) & 0.8 & 0.7 & 1.2 $\pm$ 0.8 \\
$\psi$($R_\odot$, $t_{Gal}$) ($M_{\odot}$ pc$^{-2}$ Gyr$^{-1}$) & 2.6 & 2.6 
   & 2---10  \\
$\Sigma_{gas}$($R_\odot$, $t_{Gal}$) ($M_{\odot}$ pc$^{-2}$) & 7.0 & 7.0 
   & 7---16\\
$\Sigma_{stars}$($R_\odot$, $t_{Gal}$) ($M_{\odot}$ pc$^{-2}$) & 36.3 & 36.2 
   & 35 $\pm$ 5\\
$\Sigma_{gas}$/$\Sigma_{tot}$ ($R_\odot$, $t_{Gal}$) & 0.13 & 0.13 & 
   0.05---0.20 \\
$\dot{\Sigma_{I}}$($R_\odot$, $t_{Gal}$)($M_{\odot}$ pc$^{-2}$ Gyr$^{-1}$) 
   & 1.0 & 1.0 & 0.3---1.5 \\
$\Delta$Y/$\Delta$Z & 1.9 & 1.9 & 1---3 \\
Nova outbursts (yr$^{-1}$) & 22 & 23 & 20---30 \\
X$_2(P)$/X$_2(now)$ & 1.5 & 1.5 & $<$ 3 \\
\hline
\end{tabular}
\end{table}

\newpage

\begin{table}
\begin{center}
\small
\textsc{Table 4. Solar abundances by mass ($^*$ at 4.5 Gyrs ago).}
\vspace{0.5cm}
\footnotesize
\label{}
\begin{tabular}{c c c c}
\hline
Element & $^*$A & $^a$Anders \& Grevesse (1989) & $^b$Grevesse \& Sauvaul 
(1998)\\
\hline
H & 0.71 & 0.71 & \cr
D & 3.3 ($-$5) & 4.8 ($-$5) & \cr
$^{3}$He & 2.2 ($-$5) & 2.9 ($-$5) & \cr
$^{4}$He & 2.69 ($-$1) & 2.75 ($-$1) & 2.75 ($-$1) \cr
$^{12}$C & 3.5 ($-$3) & 3.0 ($-$3) & 2.8 ($-$3) \cr
$^{16}$O & 7.1 ($-$3) & 9.6 ($-$3) & 7.7 ($-$3) \cr
$^{14}$N & 1.6 ($-$3) & 1.1 ($-$3) & 8.3 ($-$4)  \cr
$^{13}$C & 4.7 ($-$5) & 3.7 ($-$5) & \cr
$^{20}$Ne & 0.9 ($-$3) & 1.6 ($-$3) & \cr
$^{24}$Mg & 2.4 ($-$4) & 5.1 ($-$4) & \cr
Si & 6.9 ($-$4) & 7.1 ($-$4) & 7.1 ($-$4) \cr
S & 3.0 ($-$4) & 4.2 ($-$4) & 4.9 ($-$4) \cr
Ca & 3.9 ($-$5) & 6.2 ($-$5) & 6.5 ($-$5) \cr
Fe & 1.31 ($-$3) & 1.27 ($-$3) & 1.26 ($-$3) \cr
Cu & 7.7 ($-$7) & 8.4 ($-$7) & 7.3 ($-$7) \cr
Zn & 2.3 ($-$6) & 2.1 ($-$6) & 1.9 ($-$6) \cr
Z & 1.6 ($-$2) & 1.9 ($-$2) & 1.7 ($-$2) \cr
\hline
\end{tabular}\\
\end{center}
\footnotesize{$^a$ Meteoritic values.\\
$^b$ Photospheric values. The meteoritic values reported by Grevesse \& Sauval 
(1998) for Si, S, Ca, Fe, Cu, and Zn agree with the photospheric ones except 
for S, Cu, and Zn, for which they report 3.6 ($-$4), 8.7 ($-$7), and 2.2 
($-$6), respectively. The value listed here for the abundance by mass of 
$^{4}$He is that at the time of Sun formation.}
\end{table}
\begin{table}
\begin{center}
\small
\textsc{Table 5. Model parameters.}
\vspace{0.5cm}
\footnotesize
\label{}
\begin{tabular}{c c c c}
\hline
Model & $\Sigma_H(R, t_{Gal})$ & $\Sigma_{thr}^{H}$ & $\tau_H$\\
\hline
A & 17 $M_\odot$ pc$^{-2}$ for $R$ $\le$ 8 kpc & 4 $M_\odot$ pc$^{-2}$ & 0.8 
   Gyr for 4 $\le$ $R$ $\le$ 18 kpc\\
  & $\propto$ $R^{-1}$ for $R$ $>$ 8 kpc & & \\
B & 17 $M_\odot$ pc$^{-2}$ for 4 $\le$ $R$ $\le$ 18 kpc & 4 $M_\odot$ 
   pc$^{-2}$ & 0.8 Gyr for 4 $\le$ $R$ $\le$ 18 kpc\\
C & 17 $M_\odot$ pc$^{-2}$ for $R$ $\le$ 8 kpc & no threshold & 0.8 Gyr for 
   4 $\le$ $R$ $\le$ 18 kpc\\
  & $\propto$ $R^{-1}$ for $R$ $>$ 8 kpc & in the halo phase & \\
D & 17 $M_\odot$ pc$^{-2}$ for $R$ $\le$ 8 kpc & 4 $M_\odot$ pc$^{-2}$ & 
   0.8 Gyr for $R$ $\le$ 10 kpc\\
  & $\propto$ $R^{-1}$ for $R$ $>$ 8 kpc & & 2 Gyr for $R$ $>$ 10 kpc\\
\hline
\end{tabular}\\
\end{center}
\end{table}

\vfill\eject
\centerline{\bf Figure Captions}
\figcaption{ Theoretical G-dwarf metallicity distribution from Model A 
	     (our best-model) compared to the data.}
\figcaption{ a) [Fe/O] vs. [O/H] in the solar neighborhood 
            as predicted by Model A and compared to the data of Gratton et al. 
(2000). As is evident from the figure both data and model show a gap at around 
[O/H]=-0.22 dex. b) Ejected masses of O/Fe predicted by massive 
stellar evolution
models of Woosley \& Weaver 1995 (WW95) and Thieleman et al. 1996 (TNH96)
as a function of the initial stellar mass.}

\figcaption{ SFR vs. time in the solar neighborhood as predicted by Model A.}

\figcaption{ Observed radial gas density profile, togehter with predictions 
	     from Model A ({\it continuous line}), Model B ({\it long-dashed 
	     line}), Model C ({\it dotted line}), and Model D ({\it 
	     short-dashed line}) (see text for more details).}

\figcaption{ Observed radial stellar density profile, togheter with 
	     predictions from Model A ({\it continuous line}), Model B 
	     ({\it long-dashed line}), Model C ({\it dotted line}), and 
	     Model D ({\it short-dashed line}) (see text for more details).}

\figcaption{ Normalized radial SFR distribution compared with model 
             predictions: Model A ({\it continuous line}), Model B 
	     ({\it long-dashed line}), Model C ({\it dotted line}), 
	     Model D ({\it short-dashed line}) (see text for more).}

\figcaption{ Predicted radial nitrogen gradient at $t$ = $t_{Gal}$ = 14 Gyr 
	     ({\it top panel}) and $t$ = 12 Gyr ({\it bottom panel}) from 
	     Model A ({\it continuous line}), Model B ({\it long-dashed 
	     line}), Model C ({\it dotted line}), and Model D ({\it 
	     short-dashed line}) compared to observational data: Shaver 
	     et al. 1983 ({\it filled diamonds}); Fich \& Silkey 1991 
	     ({\it asterisks}); Simpson et al. 1995 ({\it filled squares}); 
	     V\' \i lchez \& Esteban 1996 ({\it open circles}); Afflerbach 
	     et al. 1997 ({\it stars}); Rudolph et al. 1997 ({\it filled 
	     circles}); Gummersbach et al. 1998 ({\it filled triangles}); 
	     Esteban et al. 1998, 1999a, b ({\it open triangles}) ({\it top 
	     panel}); Maciel \& Chiappini 1994, Maciel \& K\"oppen 1994, 
	     Maciel \& Quireza 1999 ({\it asterisks}) ({\it bottom panel}).}


\figcaption{ Predicted radial oxygen gradient at $t$ = $t_{Gal}$ = 14 Gyr 
             ({\it top panel}) and $t$ = 12 Gyr ({\it bottom panel}) from
             Model A ({\it continuous line}), Model B ({\it long-dashed
             line}), Model C ({\it dotted line}), and Model D ({\it
             short-dashed line}) compared to observational data: Shaver
             et al. 1983 ({\it filled diamonds}); Fich \& Silkey 1991
             ({\it asterisks}); Simpson et al. 1995 ({\it filled squares});
             V\' \i lchez \& Esteban 1996 ({\it open circles}); Afflerbach
             et al. 1997 ({\it stars}); Rudolph et al. 1997 ({\it filled
             circles}); Smartt \& Rolleston 1997 ({\it open squares});
             Gummersbach et al. 1998 ({\it filled triangles});
             Esteban et al. 1998, 1999a, b ({\it open triangles});
             Deharveng et al. 2000 ({\it open diamonds}) ({\it top
             panel}); Maciel \& Chiappini 1994, Maciel \& K\"oppen 1994,
             Maciel \& Quireza 1999 ({\it asterisks}) ({\it bottom panel}).}

\figcaption{ Predicted radial sulfur gradient at $t$ = $t_{Gal}$ = 14 Gyr 
	     ({\it top panel}) and $t$ = 12 Gyr ({\it bottom panel}) from 
	     Model A ({\it continuous line}), Model B ({\it long-dashed 
	     line}), Model C ({\it dotted line}), and Model D ({\it 
	     short-dashed line}) compared to observational data: Shaver 
	     et al. 1983 ({\it filled diamonds}); Simpson et al. 1995 
	     ({\it filled squares}); V\' \i lchez \& Esteban 1996 ({\it 
    	     open circles}); Afflerbach et al. 1997 ({\it stars}); Rudolph 
	     et al. 1997 ({\it filled circles}) ({\it top panel}); Maciel 
	     \& Chiappini 1994, Maciel \& K\"oppen 1994, Maciel \& Quireza 
	     1999 ({\it asterisks}) ({\it bottom panel}).}

\figcaption{ Predicted radial iron gradient at $t$ = $t_{Gal}$ = 14 Gyr 
	     from Model A ({\it continuous line}), Model B ({\it long-dashed 
	     line}), Model C ({\it dotted line}), and Model D ({\it 
	     short-dashed line}) compared to observational data: Twarog et al. 
	     1997 ({\it stars}); Carraro et al. 1998 ({\it filled pentagons}); 
	     Friel 1999 ({\it filled triangles}).}

\figcaption{ Predicted radial gradients for N, O, S and Fe at $t$ = 2 
             ({\it long-dashed line}), 5 ({\it dashed line}), 9.5 
             ({\it dotted line}) and 14 ({\it solid line}) Gyrs for Models 
	     A to D (from the top to bottom).}


\figcaption{ Predicted evolution of the abundance gradients for N, O, S,
             and Fe between 4\,--\,14 kpc from the centre, for Models A to D
	     (from top to bottom).}

\figcaption{ Predicted $<[O/Fe]>$ ratios for the stars at any Galactocentric 
distance, averaged by mass, i.e. they correspond to the [O/Fe] ratio of the 
stellar population dominating in mass,
as predicted by Model A. Panel a): $<[O/Fe]>$ vs. R,
i.e. the ratios versus the Galactocentric distance corresponding to the stellar 
birthplace; Panel b) $<[O/Fe]>$ vs. Age, where Age refers to the age of the
dominant stellat population. From both figures it appears that there are clear 
correlations between the abundance ratios and birthplace and age.}

\newpage

\begin{figure}
\figurenum{1}
\centerline{\psfig{figure=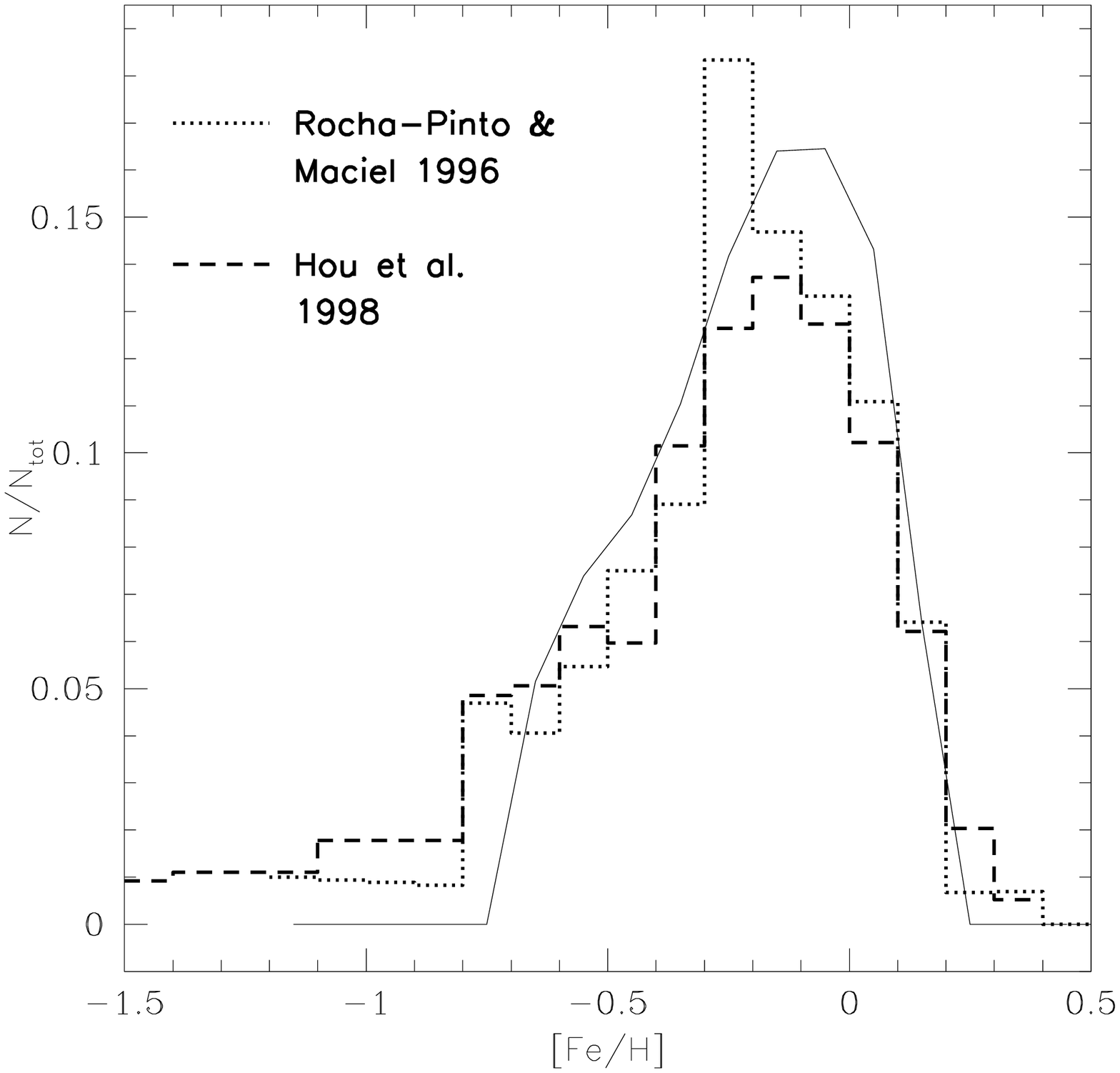,width=14cm,height=14cm} }
\caption{}
\end{figure}

\newpage

\begin{figure}[ht]
\figurenum{2a}
\centerline{\psfig{figure=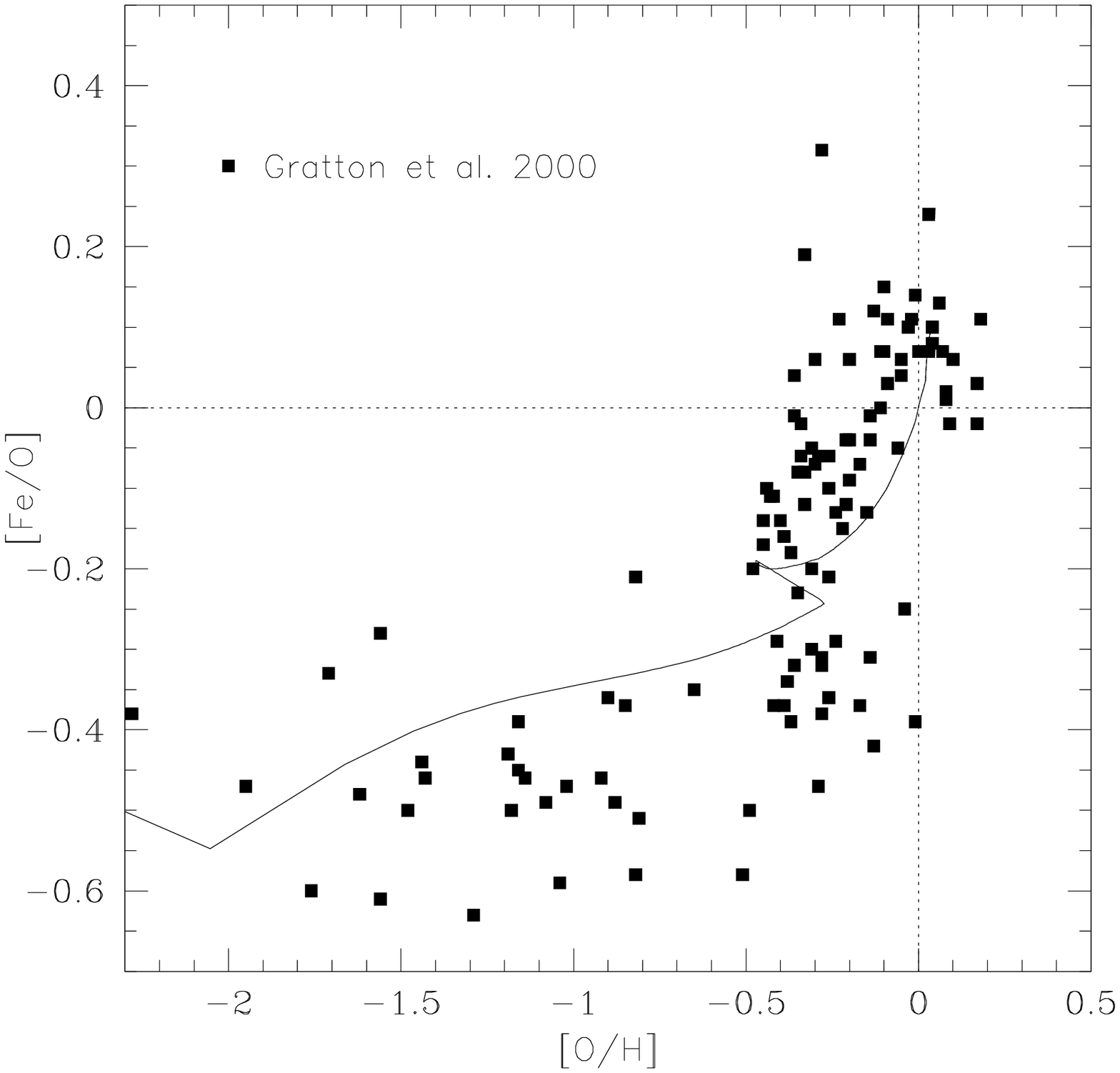,width=14cm,height=14cm} }
\caption{}
\end{figure}

\newpage

\begin{figure}[ht]
\figurenum{2b}
\centerline{\psfig{figure=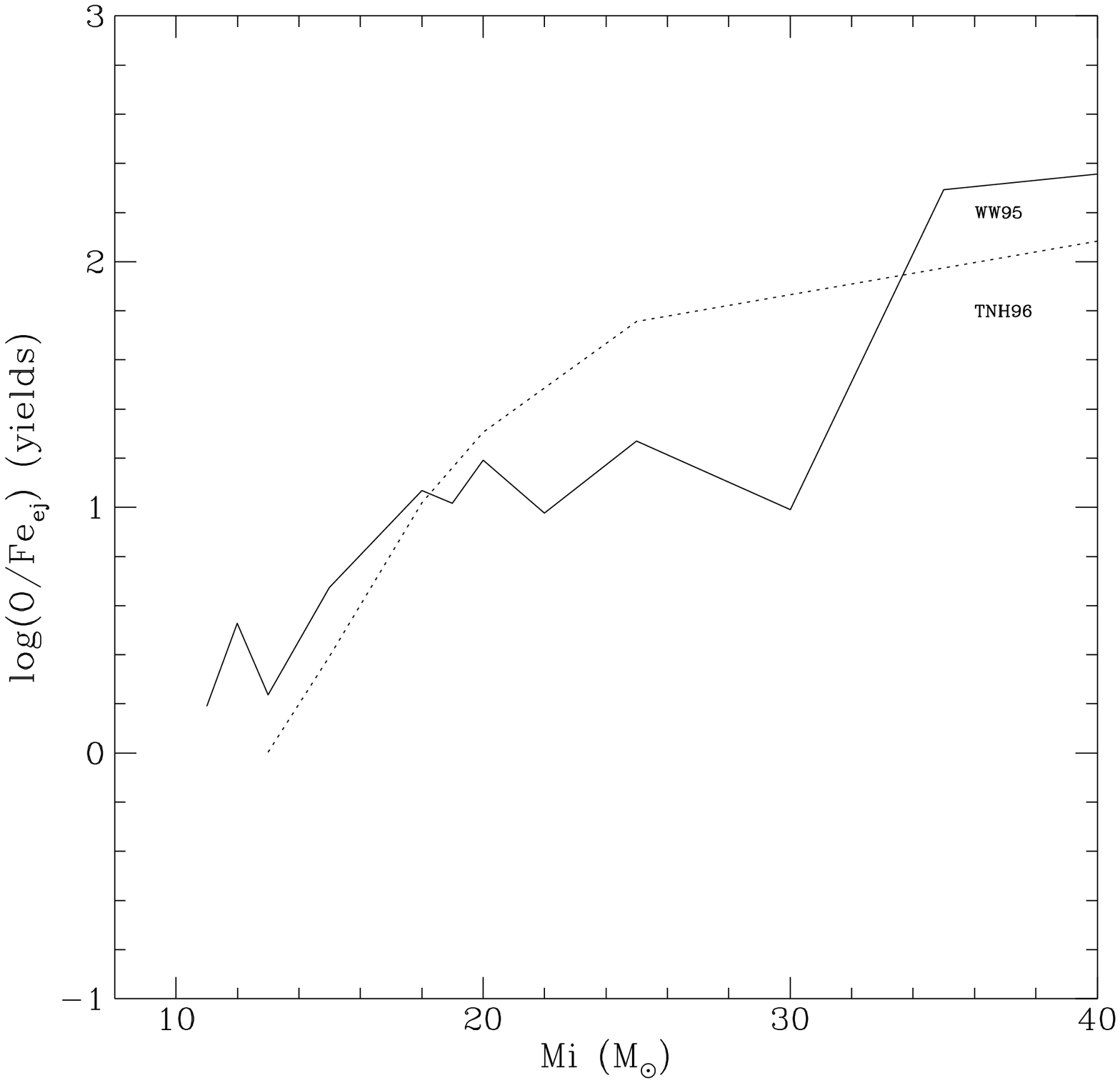,width=14cm,height=14cm} }
\caption{}
\end{figure}

\newpage

\begin{figure}[ht]
\figurenum{3}
\centerline{\psfig{figure=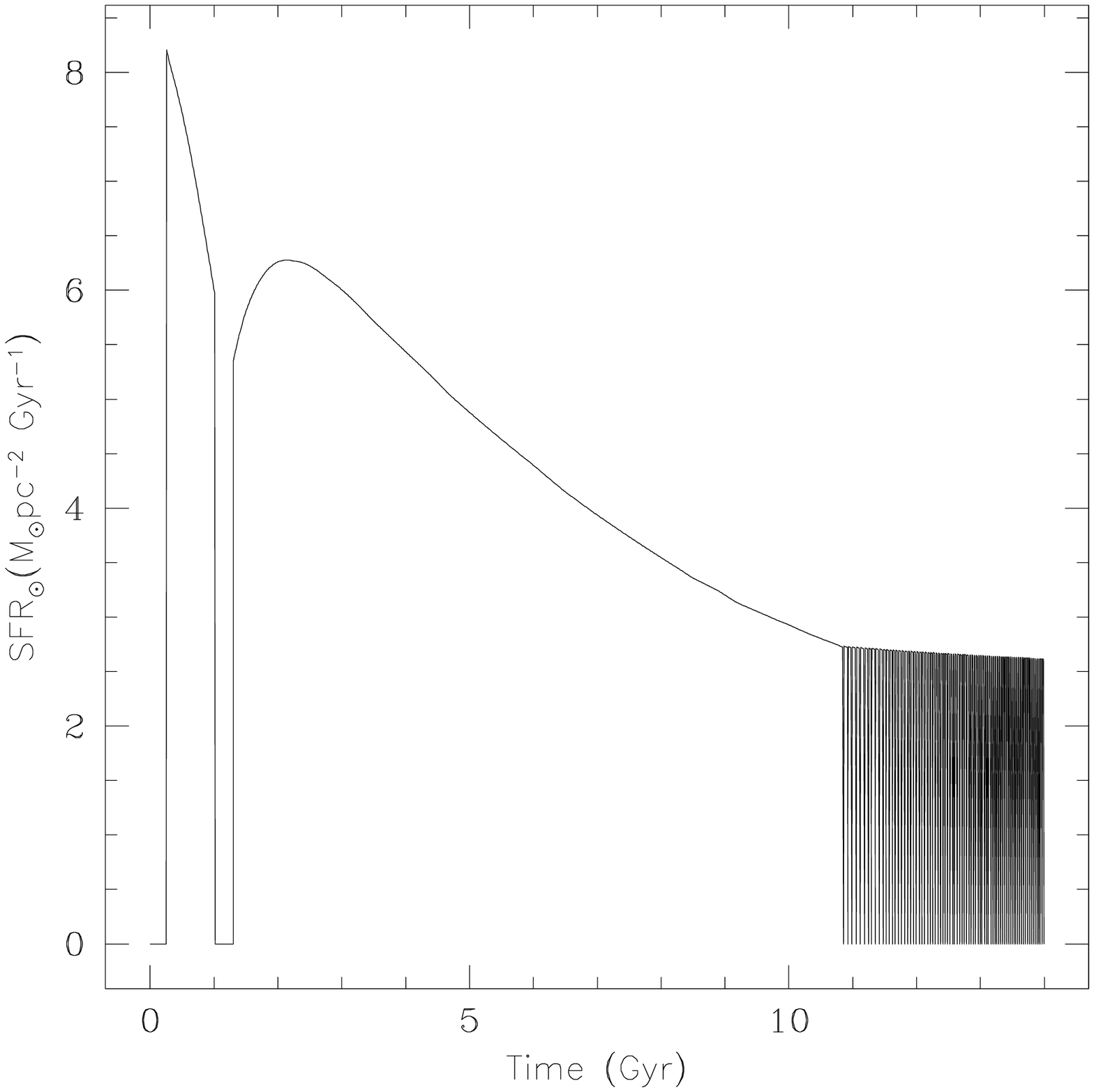,width=14cm,height=14cm} }
\caption{}
\end{figure}

\newpage

\begin{figure}[ht]
\figurenum{4}
\centerline{\psfig{figure=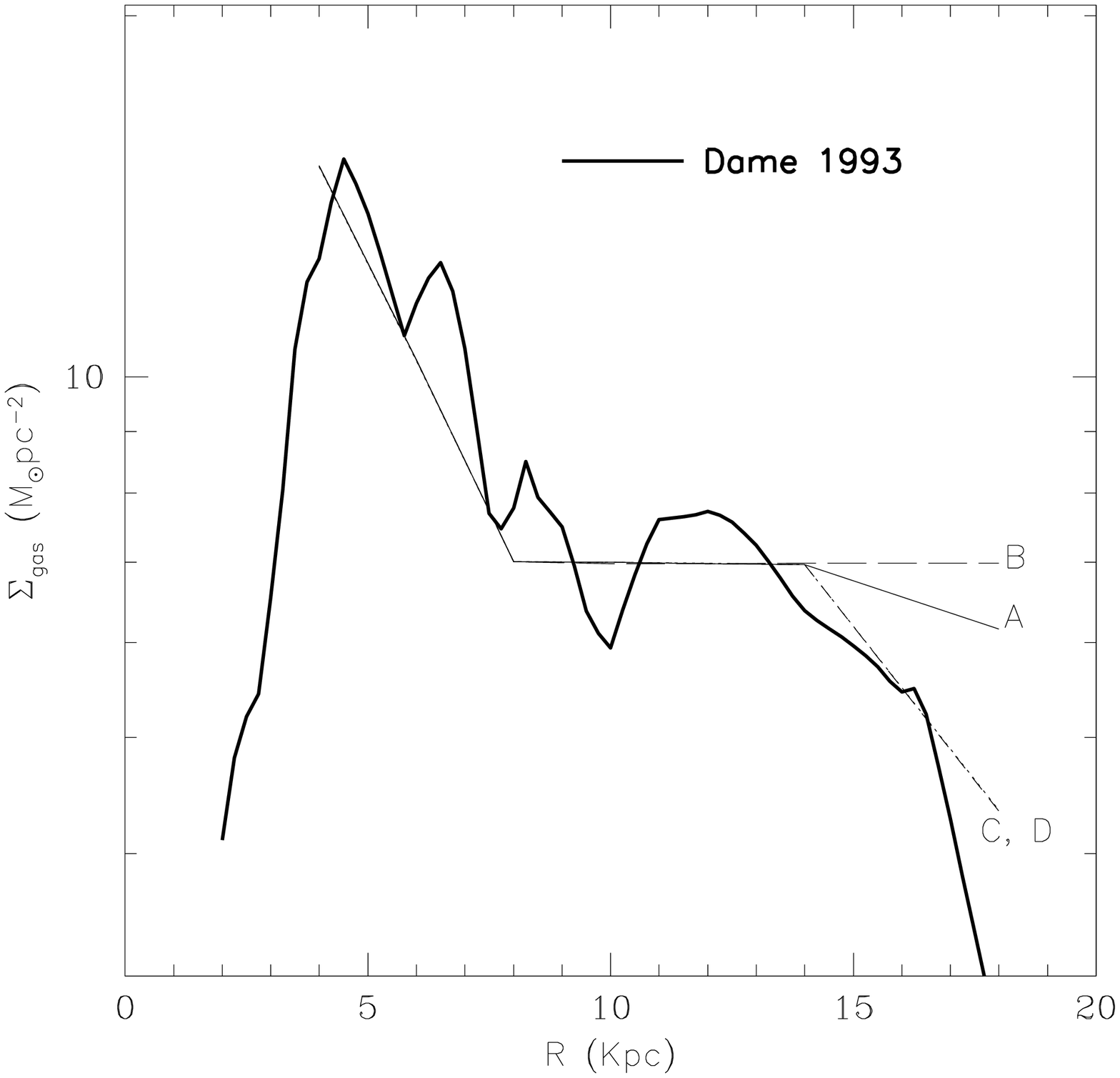,width=14cm,height=14cm} }
\caption{}
\end{figure}

\newpage

\begin{figure}[ht]
\figurenum{5}
\centerline{\psfig{figure=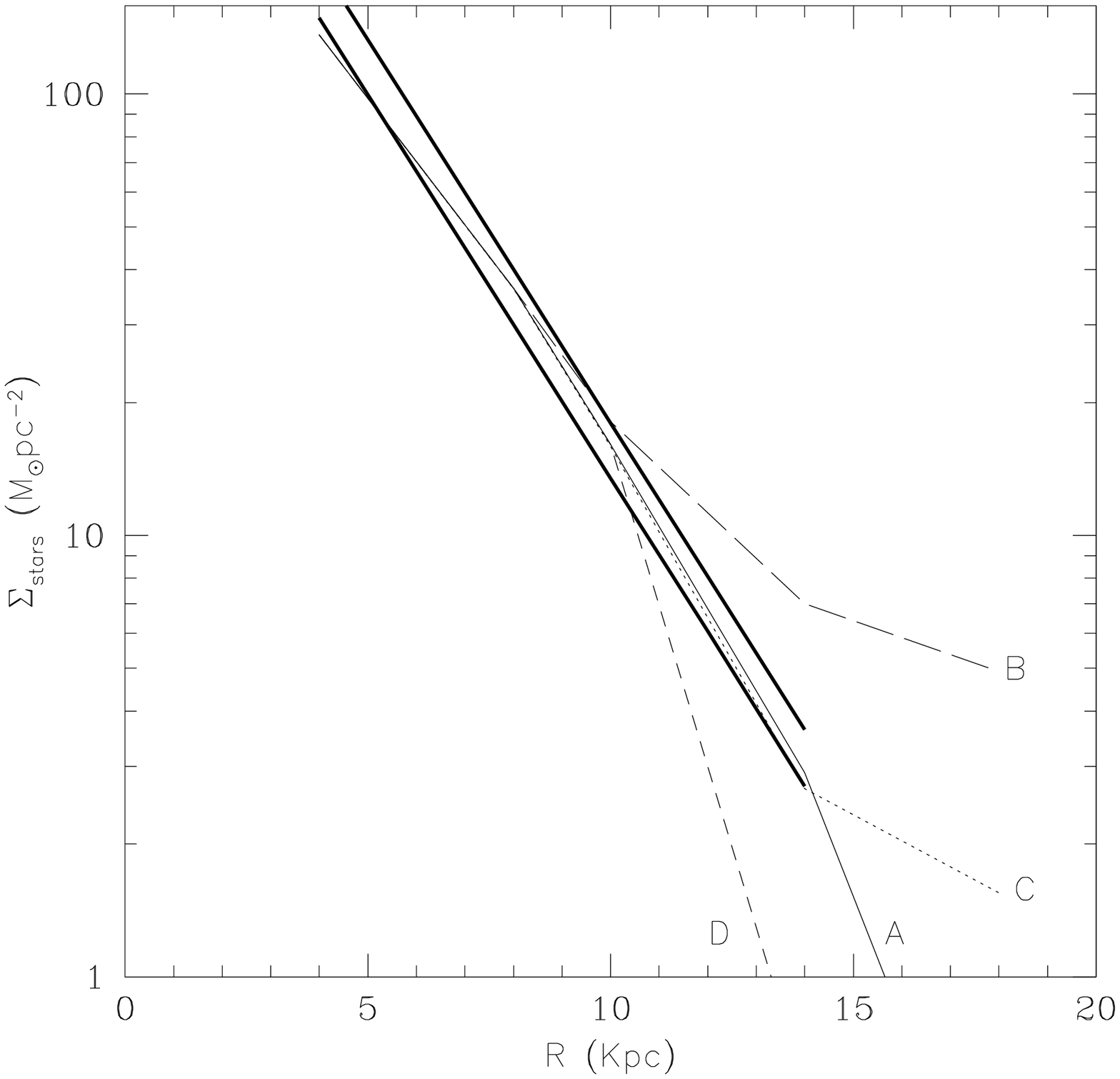,width=14cm,height=14cm} }
\caption{}
\end{figure}

\newpage

\begin{figure}[ht]
\figurenum{6}
\centerline{\psfig{figure=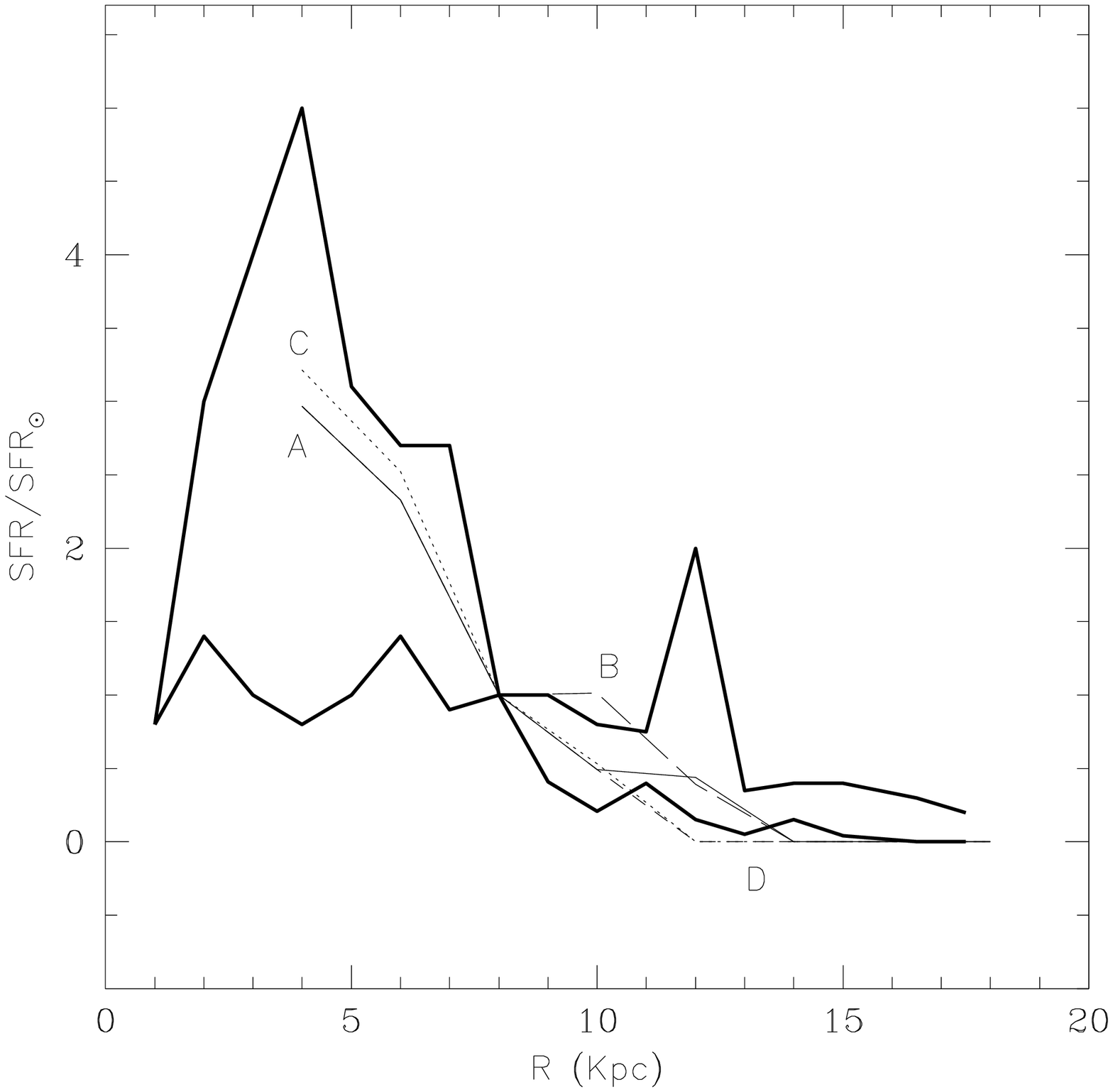,width=14cm,height=14cm} }
\caption{}
\end{figure}

\newpage

\begin{figure}[ht]
\figurenum{7}
\centerline{\psfig{figure=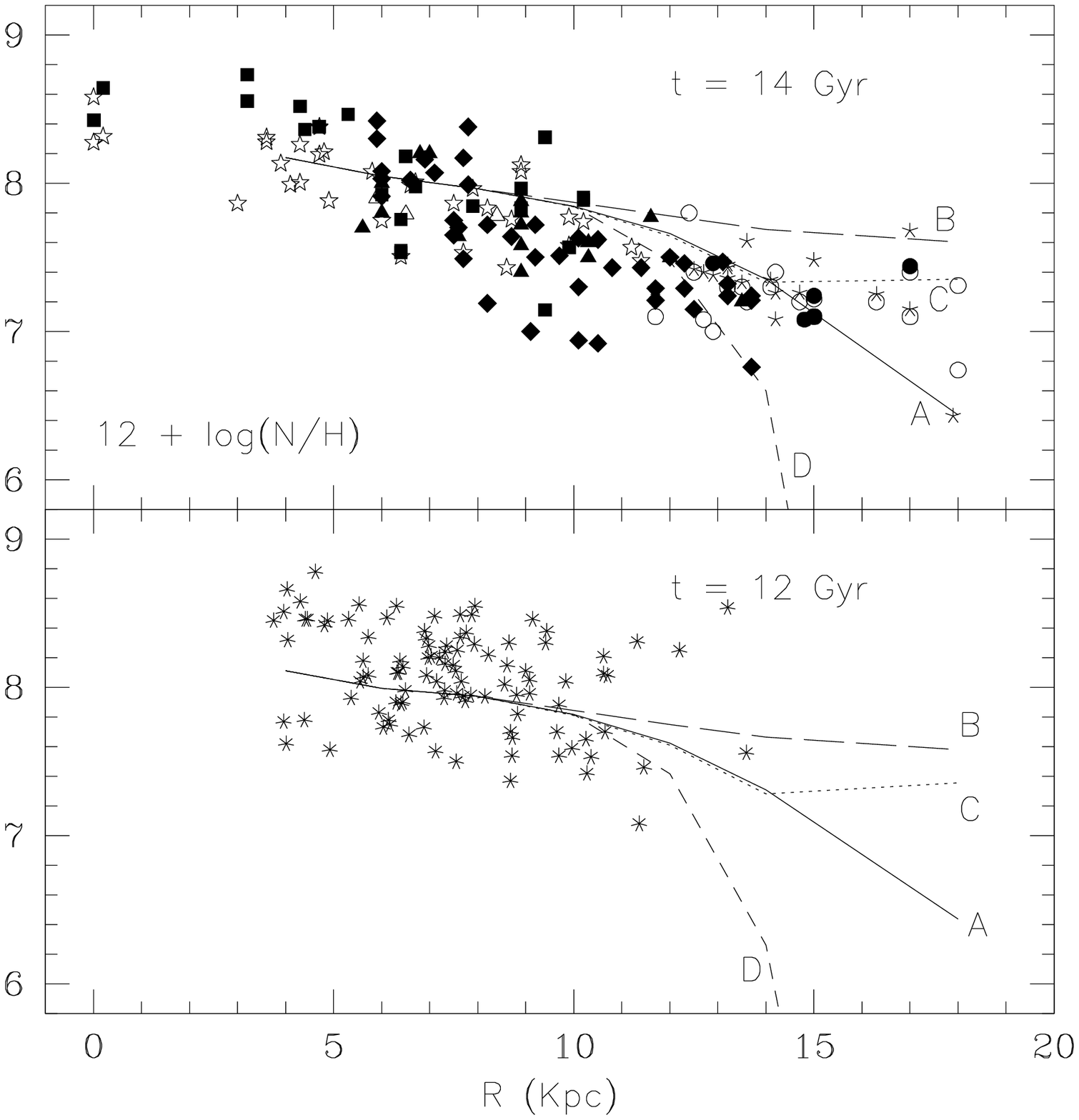,width=14cm,height=14cm} }
\caption{}
\end{figure}

\newpage

\begin{figure}[ht]
\figurenum{8}
\centerline{\psfig{figure=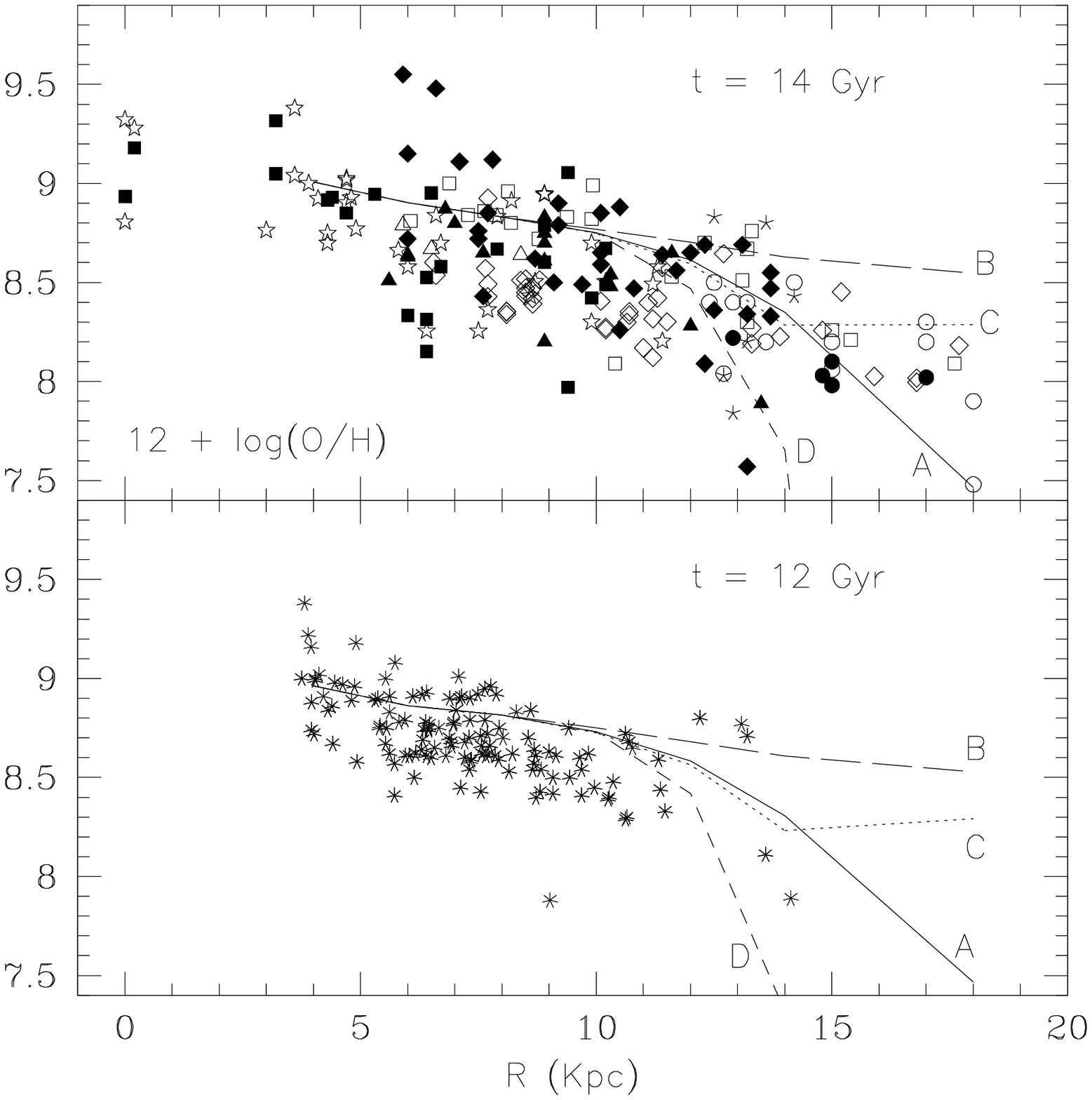,width=14cm,height=14cm} }
\caption{}
\end{figure}

\newpage

\begin{figure}[ht]
\figurenum{9}
\centerline{\psfig{figure=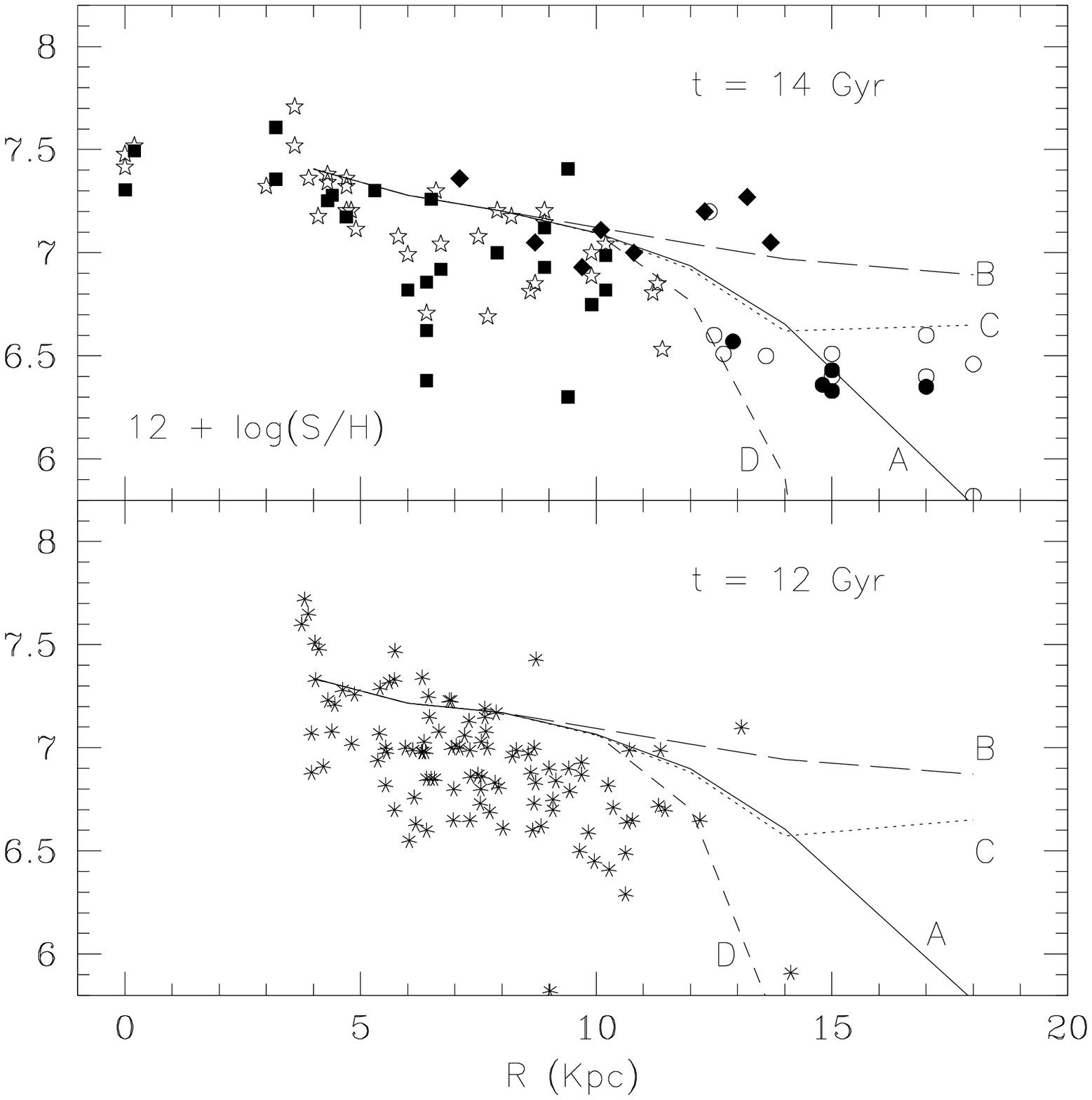,width=14cm,height=14cm} }
\caption{}
\end{figure}

\newpage

\begin{figure}[ht]
\figurenum{10}
\centerline{\psfig{figure=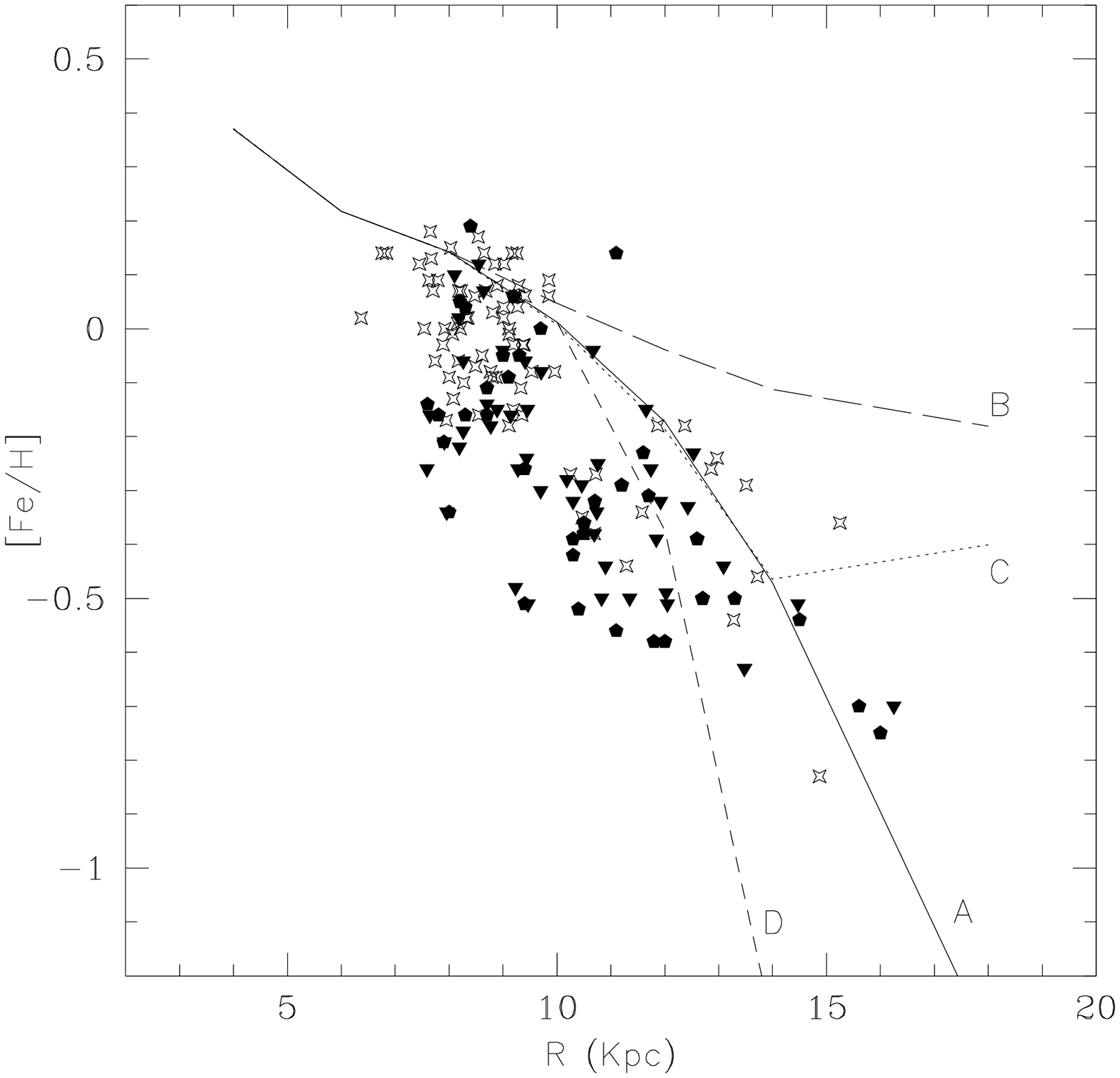,width=14cm,height=14cm} }
\caption{}
\end{figure}

\newpage

\begin{figure}[ht]
\figurenum{11}
\centerline{\psfig{figure=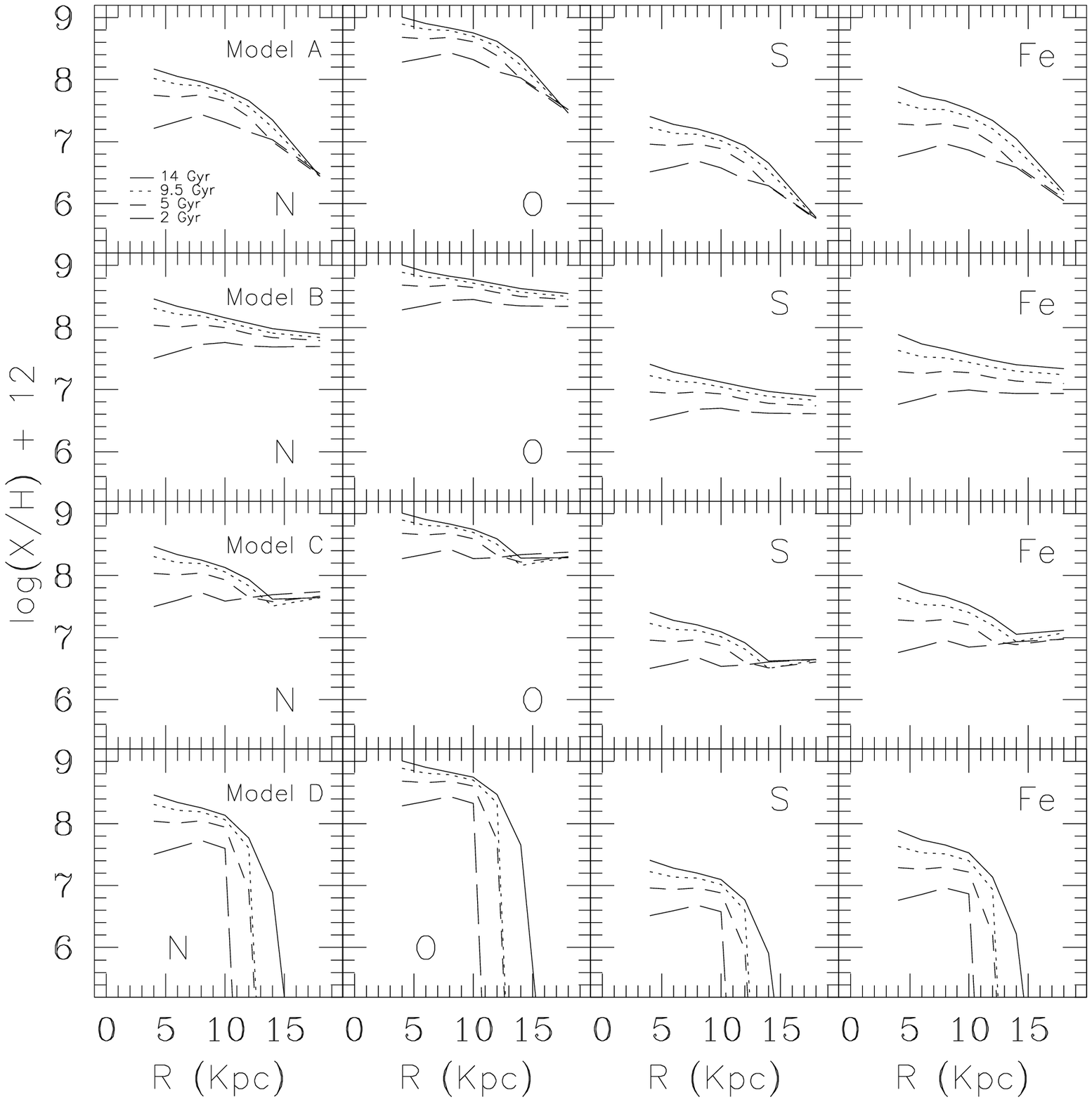,width=14cm,height=14cm} }
\caption{}
\end{figure}

\newpage


\newpage

\begin{figure}[ht]
\figurenum{12}
\centerline{\psfig{figure=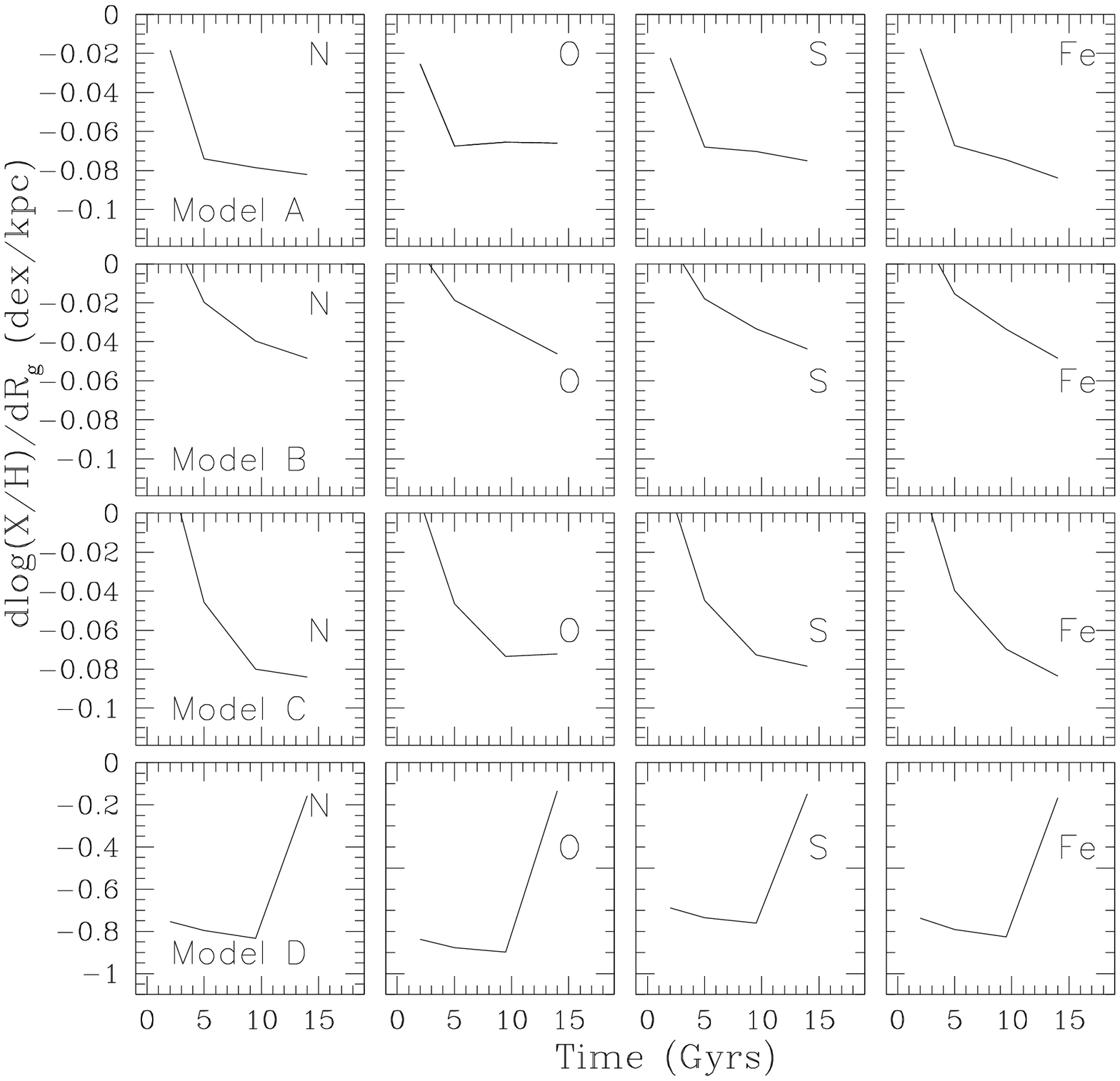,width=14cm,height=14cm} }
\caption{}
\end{figure}

\newpage

\begin{figure}[ht]
\figurenum{13}
\centerline{\psfig{figure=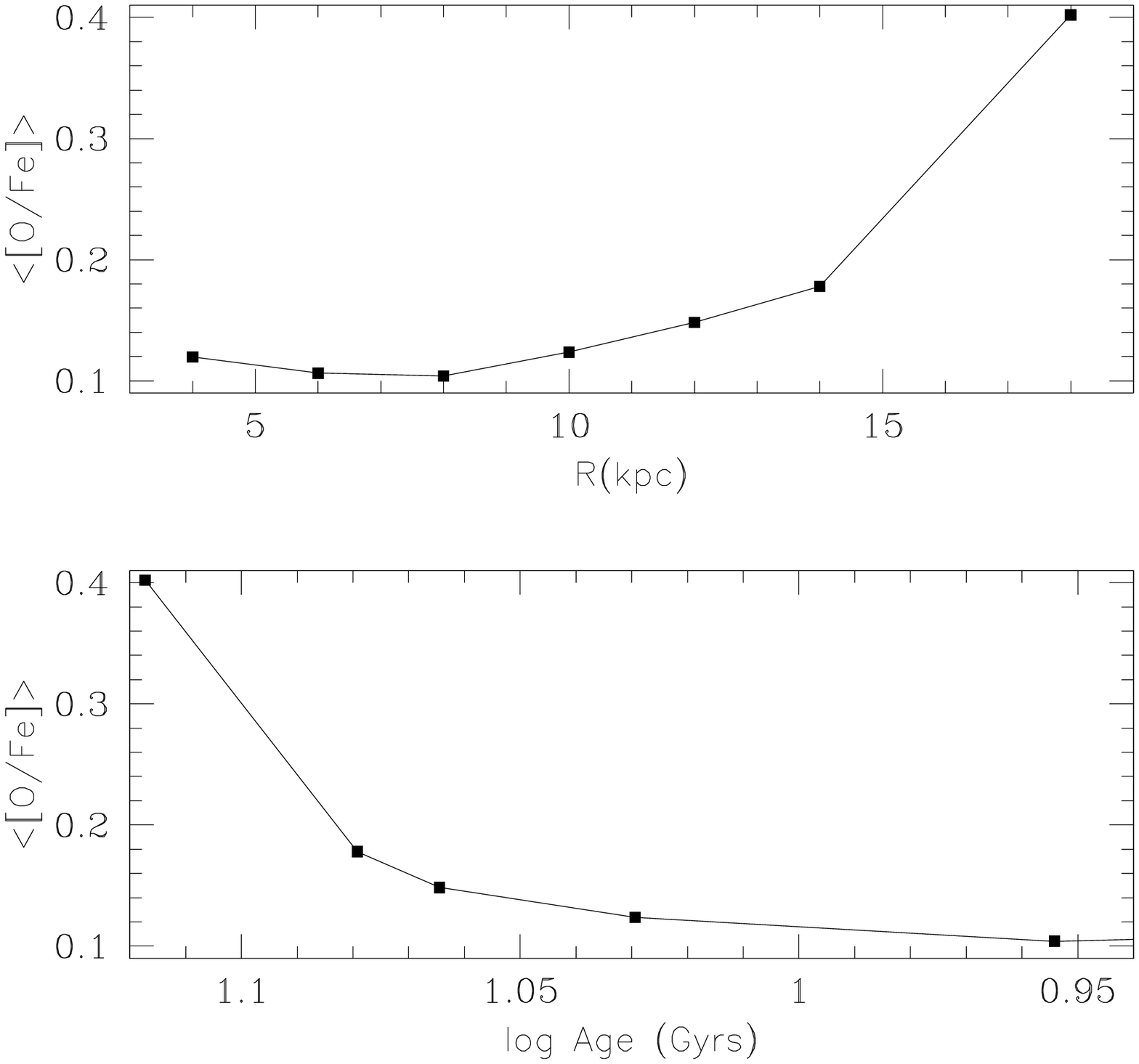,width=14cm,height=14cm} }
\caption{}
\end{figure}

\end{document}